\documentclass[pra,twocolumn,superscriptaddress,showpacs,amsmath,amssymb,showpacs]{revtex4}

\pacs{%
03.75.Kk, %Bose-Einstein condensation dynamic properties
05.70.Jk, %QCP
37.25.+k, %Atom interferommetry
74.40.Gh %Nonequilibrium fluctuations
}

\usepackage{latexsym}
\usepackage{amsmath, amsthm, amssymb} 
\usepackage{txfonts}
\usepackage{natbib,hyperref}%to get nice hyperlinks in the references

\usepackage{graphicx}% Include figure files
\usepackage{dcolumn}% Align table columns on decimal point
\usepackage{bm}
\usepackage{txfonts}
\usepackage{mathrsfs}
\usepackage{feynmf}
\usepackage{comment}

\usepackage{float}

\usepackage[latin1]{inputenc}
\usepackage{amsmath}
\usepackage{amsfonts}
\usepackage{amssymb}
\usepackage{setspace}
\usepackage[usenames]{color}
\usepackage{array}
\def  \citeSM{{(see Appendix)}}

\usepackage[latin1]{inputenc}
\usepackage{amsmath}
\usepackage{amsfonts}
\usepackage{amssymb}
\usepackage{graphicx}
\usepackage{setspace}
\usepackage[usenames]{color}
\usepackage{epstopdf}
\usepackage{array}
\newcommand{\be}{\begin{equation}}
\newcommand{\ee}{\end{equation}}

\def \be{\begin{equation}}
\def \ee{\end{equation}}
\def \ba{\begin{array}}
\def \ea{\end{array}}
\def \bea{\begin{eqnarray}}
\def \eea{\end{eqnarray}}
\def \nn{\nonumber}
\def \e{{\epsilon}}

\def \a{{\alpha}}

\def \D{{\Delta}}

\def \w{{\omega}}

\def \e{{\epsilon}}

\def \yd{^\dagger}
\def \av#1{{\langle#1\rangle}}
\def \ket#1{{\,|\,#1\,\rangle\,}}

\def \braket#1#2{{\,\langle\,#1\,|\,#2\,\rangle\,}}

\def \beas{\begin{eqnarray*}}
\def \eeas{\end{eqnarray*}}

\newcounter{indice}

\def \bn{\begin{enumerate}}
\def \en{\end{enumerate}}

\def \bb{\begin{thebibliography}{99}}
\def \eb{\end{thebibliography}}

\def\coloronline{{[Color online] }}

\begin{document}
\title{Universal Rephasing Dynamics\\ after a Quantum Quench via Sudden Coupling of Two Initially Independent Condensates}

\author{Emanuele G. Dalla Torre}
\affiliation{Department of Physics, Harvard University, Cambridge MA 02138}

\author{Eugene Demler}
\affiliation{Department of Physics, Harvard University, Cambridge MA 02138}

\author{Anatoli Polkovnikov}
\affiliation{Department of Physics, Boston University, Boston MA 02215}

\begin{abstract}
We consider a quantum quench in which two initially independent condensates are suddenly coupled, and study the subsequent ``rephasing'' dynamics. For weak couplings, the time-evolution of physical observables is predicted to follow universal scaling laws, connecting the short-time dynamics to the long-time non-perturbative regime. We first present a two-mode model valid in two and three dimensions and then move to one dimension, where the problem is described by a gapped Sine-Gordon theory. Combining analytical and numerical methods, we compute universal time-dependent expectation values, allowing a quantitative comparison with future experiments.
\end{abstract}

\maketitle

%\section{Introduction}

Equilibrium systems near second-order phase transitions are the best known examples of universal behavior in many-body physics. For these systems, universality implies that all macroscopic quantities are related to the distance from the transition point through universal critical exponents.  In recent years, many efforts have been dedicated to the extension of these concepts to quantum systems out of thermal equilibrium (see for example Refs.~\cite{mitra06,diehl08,dallatorre10} for open systems and Ref.~\cite{polkovnikov11} for a review on closed systems). In particular, some authors considered closed systems that are slowly driven through a second-order phase transition and, extending the Kibble-Zurek argument, found universal scaling laws of static and dynamic quantities \cite{deng08,degrandi11,chandran12,kolodrubetz12}. Others considered sudden quenches and predicted a universal scaling of physical observables at long times after the quench \cite{degrandi10A,campos10,yier12,karrasch12}. Here we extend these arguments by demonstrating a universal behavior in the transient dynamics following a sudden quench, both at short and long times.

\begin{figure}[t]
\includegraphics[scale=0.75]{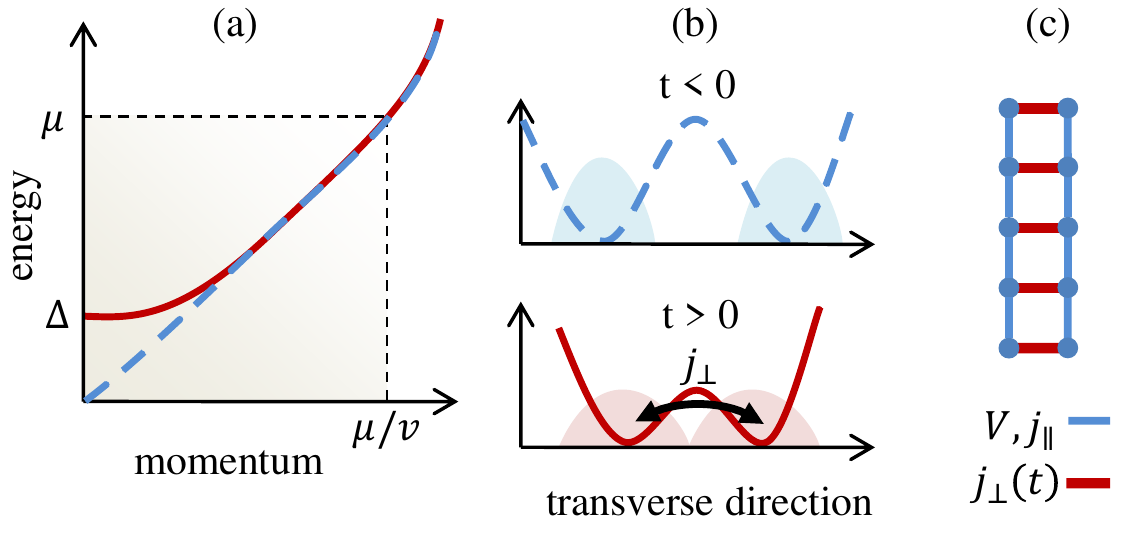}\vspace{-0.2cm}
\caption{\coloronline (a) Dispersion relation before (dashed) and after (solid) a quantum quench opening a gap $\Delta$ and involving only the low-frequency modes. (b) Physical realization: sudden coupling of two independent condensates via a uniform tunneling $j_\perp$. (c) Lattice model used for the numerical calculations.}\label{fig:schematic}
\end{figure}

In this paper we consider systems that are initially prepared in the ground state of a gapless Hamiltonian, and that are quenched by the sudden opening of an excitation gap $\D$ (see Fig.~\ref{fig:schematic}(a)). In the Renormalization Group (RG) language, this corresponds to the sudden switching-on of a  ``relevant'' perturbation. The resulting dynamics is described by a strongly interacting theory and is intrinsically non-perturbative \cite{clemens12}.  Experimentally, this can be realized for example by preparing two independent Bose-Einstein condensates in their respective ground states and then suddenly switching-on a finite tunneling coupling $j_\perp$ (see Fig.~\ref{fig:schematic}(b)). This opens an excitation gap for the relative-phase modes, given at a mean field level by the plasma frequency $\Delta=2\sqrt{\mu j_\perp}$, where $\mu$ is the chemical potential. Beyond-mean-field effects will be extensively discussed below.

In the limit of $j_\perp\ll\mu$, the dynamics following the quench is dominated by the low-frequency modes and is thus expected to be universal. For a generic physical observable $C(t)$ we predict the universal contributions to have the form
\be C(t) = \left(\frac{\D}{\mu}\right)^\eta R\left(\D t\right)\;.\label{eq:scalinglaw}\ee
Here $\eta$ is the scaling dimension corresponding to the specific observable and $R$ is a scaling function, fixing the universal behavior of the transient many-body dynamics. As we will demonstrate, the scaling ansatz (\ref{eq:scalinglaw}) is valid at times both shorter and longer than the inverse gap $1/\Delta$. This allows to extract information about the non-perturbative long-time dynamics from the short-time dynamics, where time-dependent perturbation theory applies. A similar relation is known to hold for classical systems, with important theoretical and computational applications \cite{janssen89,zheng98,kolton06,albano11}.

%For the sake of clarity througout this Letter we focus on a single physical observable, the intereference contrast $C_{12}(t)$, leaving the analysis of the universal contribution to other observables for a later study. For a homogeneus system with average density $\rho_0$ the contrast is given by
%\be C_{12}(t) =\frac1{2\rho_0}\av{\psi\yd_1(x,t)\psi\nd_2(x,t)+{\rm H.c.}},\ee 
%where $\psi_1(x)$ and $\psi_2(x)$ are complex bosonic fields describing the two condensates. In the initial state the two condensates are independent and the relative phase varies from shot to shot, leading to interference fringes (``matter waves'') at random position \cite{cirac96,andrews97}. Averaging over several realizations, one obtains a flat distribution with zero contrast $C_{12}(t=0)=0$. When the coupling $j_\perp$ is turned on, atoms can tunnel from one condensate to the other, leading to a ``rephasing'' of the two condensates which we will show to be universal and follow eq. (\ref{eq:scalinglaw}).

\begin{figure}[t]
\includegraphics[scale=0.8]{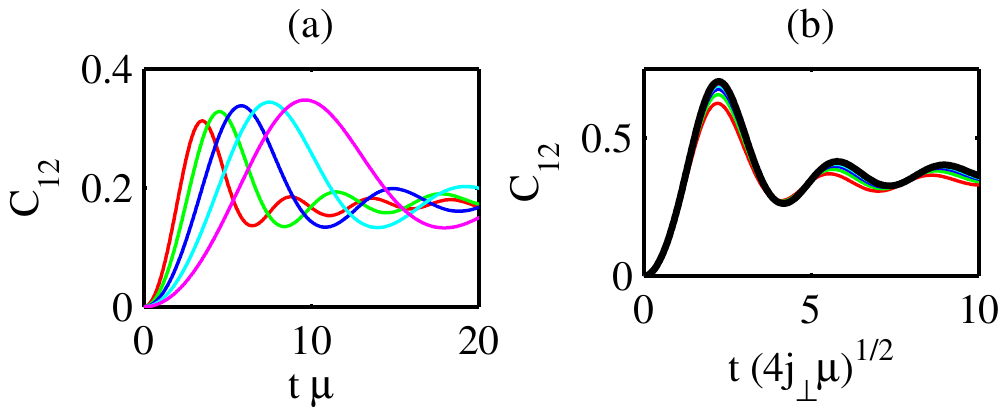}\vspace{-0.3cm}
\caption{\coloronline Time evolution of the interference contrast $C_{12}(t)=\av{\psi_1\yd\psi_2+{\rm H.c.}}/2N = \av{\cos(\sqrt{2}\phi)}$, according to the two-mode quantum Hamiltonian (\ref{eq:HMF}) with $N=1000$. (a) Exact diagonalization for different values of the ratio $j_\perp/\mu$ = 0.10 {\rm (red)},~0.060 {\rm (green)},0.037 {\rm (blue)},~0.022 {\rm (cyan)},~0.0135 {\rm (magenta)}. (b) Same plots on rescaled axes, showing the universal behavior (\ref{eq:scalinglaw}) with $\eta=0$, superimposed to the analytic solution of the simple pendulum (thick solid line).}\vspace{-0.3cm}
\label{fig:meanfield}
\end{figure}

{\it Two and three dimensions --} 
To substantiate our predictions we first apply a two-mode model, valid in two and three dimensions \footnote{A homogeneous tunneling coupling in three dimensions can be realized using two different hyperfine states and suddenly turning-on a frequency-matched microwave coupling.}. Further conditions for the validity of this model are discussed towards the end of the paper. 
%This approach will demonstrate the main features appearing in the more complicate, one dimensional case, to be studied below. 
By considering only the macroscopically occupied state of each condensate, we describe the sudden quench by:
\be H(t) = \frac{\mu}{N}\left(\delta n_1^2 + \delta n_2^2\right) - \Theta(t) j_\perp \left(\psi\yd_1\psi_2 +{\rm H.c.}\right)\label{eq:HMF}\;.\ee
Here $\delta n_\a=\psi\yd_\a\psi_\a-N_\a$, with $N_\a=N/2$; $\psi_\a$, with $\a=1,2$, is a canonical bosonic operators, and $\Theta(t)$ is the Heaviside step function. The two terms in Eq.~(\ref{eq:HMF}) describe respectively the interactions among the atoms in the same condensate and the sudden switching-on of a tunneling coupling. Results from the exact diagonalization of Eq.~(\ref{eq:HMF}) for different values of $j_\perp$ are shown in Fig.~\ref{fig:meanfield}(a). Subplot (b) shows the data collapse obtained by rescaling the time axis, and demonstrating the validity of our scaling ansatz (\ref{eq:scalinglaw}), with $\eta=0$.

%Exploiting the conservation of the total number of atoms $N = N_1+N_2 = \psi\yd_1\psi_1 + \psi\yd_2\psi_2$, we can express the Hamiltonian (\ref{eq:HMF}) in terms of a single set of operators, $S_z = (N_1-N_2)/2$ and $S^+ = \psi_1\yd\psi_2$, satisfying the SU(2) spin algebra. In this language eq. (\ref{eq:HMF})  acquires a more compact form:
%\be H(t) = 2\frac{\mu}{N} S_z^2 + 2\Theta(t)~ j_\perp S_x \label{eq:Hspin},\ee
%where $S_x = (S^++S^-)/2$. To complete the mapping we note that the inital state corresponds to the state $\ket{S=N/2,S_z=0}$ and that $C_{12}(t) = \av{S_x}/N$. 

To understand the observed scaling behavior, it is useful to introduce the phase fields $\phi_\a$, canonically conjugate to $\delta n_\a$, through the approximate relation $\psi_\a\approx \sqrt{N_\a}e^{i\phi_\a}$. Moving to the relative coordinates $\phi=(\phi_1-\phi_2)/\sqrt{2}$ and $n = (\delta n_1-\delta n_2)/\sqrt2$, we obtain the Josephson junction Hamiltonian
\be H(t) = \frac{2\mu}{N}n^2 - 2\Theta(t)~ j_\perp N \cos(\sqrt2\phi)\label{eq:HMF2}\;.\ee
The ground state and first-few excited states of the final Hamiltonian $H(t>0)$ are well described by the harmonic approximation $\cos(\sqrt{2}\phi)\approx 1- \phi^2$. When applied to the present quench, this approximation predicts undamped oscillations with frequency $\D=2\sqrt{\mu j_\perp}$. Instead, as shown in Fig.~\ref{fig:meanfield}, the universal dynamics displays strongly damped oscillations%, with a damping coefficient of order one. 
. The harmonic approximation breaks down because the initial state $\ket{n=0}$ has a large overlap with a macroscopic number of eigenstates of the final Hamiltonian, $N_{\rm excited} = N \sqrt{j_\perp/\mu} \gg 1$. For these states all higher-order Taylor components of the cosine need to be considered, highlighting the strongly interacting nature of the predicted dynamics. 

In the limit of large $N$, an alternative, non-perturbative description is available: the semiclassical approach, or Truncated Wigner approximation \cite{steel98,sinatra01,bodet10}. Following this method \citeSM, we map the quantum Hamiltonian (\ref{eq:HMF2}) to the classical equations of motion of a  simple pendulum $d^2\phi/dt^2 = - \D^2 \sin(\sqrt{2}\phi)/\sqrt{2}$, whose analytical solution is known. The quantum nature of the problem enters through the initial conditions, for which $\phi$ is uniformly distributed between zero and $\sqrt{2}\pi$. The resulting dynamics corresponds to damped oscillations with a single frequency scale $\D$ and correctly reproduces the numerical solution of the full quantum model, as shown in Fig.~\ref{fig:meanfield}(b).

{\em One dimension --} In one dimension the initial state is not described by two macroscopically occupied modes, but rather by two ``quasi-condensates'' with power-law correlations. 
%As we will see, this does not prevent the {\it relative} phase to acquire long-range order, inducing a finite contrast of the average interference fringes. 
Using the standard bosonization technique \cite{giamarchi_book}, we describe the system in terms of two phase {\it fields}, $\phi_{1}(x,t)$ and $\phi_2(x,t)$, and the time-dependent Hamiltonian
\be H = \sum_{\a=1,2}H_{LL}\left[\phi_\a\right] - 2j_\perp(t)~\rho_0\int dx~\cos(\phi_1-\phi_2). \label{eq:HSG}\ee
Here the first term describes the low-frequency modes of the two independent condensates as ``Tomonaga-Luttinger liquids'' \cite{haldane80} with sound velocity $v$ and Luttinger parameter $K$. The latter measures the ratio between the  longitudinal kinetic energy and the interaction energy: $K\to \infty$ corresponds to free bosons and $K=1$ to the Tonks-Girardeau limit. The second term of Eq.~(\ref{eq:HSG}) describes the tunneling between the two condensates, according to the ``bosonization dictionary'' $\psi_{\a}\to\sqrt{\rho_0}e^{i\phi_{\a}}$. 

Introducing the relative coordinate $\phi=(\phi_1-\phi_2)/\sqrt{2}$ we obtain the well-known ``Sine-Gordon'' model with a non-linear term proportional to the cosine of the relative phase, $\cos(\phi_1-\phi_2)=\cos(\sqrt{2}\phi)$. At zero temperature, this model is characterized by a quantum phase transition between the gapless Tomonaga-Luttinger liquid ($K<1/4$) and a gapped phase ($K>1/4$).
%
%phi_old=phi_new/sqrt(2)
%K_old=2/K_new$
%
The effects of the quench in the two phases are very different. For $K<1/4$, the cosine is irrelevant in an RG sense and the low-frequency dynamics is governed by the quadratic Luttinger liquid theory (i.e. the first term of the Hamiltonian (\ref{eq:HSG})). At long times, this theory predicts the appearance of universal power-laws \cite{calabrese06,karrasch12}. In contrast, for $K>1/4$, the cosine is relevant and the resulting theory is strongly interacting \footnote{The dynamics in the gapped phase has been studied in Ref.~\cite{iucci10} by a quadratic expansion of the cosine term. This approach neglects the mode-coupling effect of the cosine and does not capture the rephasing dynamics.}. As we will now show, in this case the dynamics is universal both at short and long times.

To study the dynamics of the problem we first perform a series expansion in the tunneling coupling $j_\perp$. Moving to a path-integral formulation of the problem (for an introduction see for example Refs.~\cite{altland_book,kamenev_book}), we represent the Hamiltonian (\ref{eq:HSG}) is terms of the real-time action:
\be S = \int_{\gamma_K}dt\int dx~\frac{K}{\pi}\left[(\partial_x\phi^2)-(\partial_t\phi^2)\right] + 2\Theta(t)~ j_\perp\rho_0 \cos(\sqrt{2}\phi)\label{eq:SSG}\;. \ee
Here $\gamma_K$ is the Keldysh contour and we switched to units where $v=1$. Splitting the Keldysh contour in its forward and backward branches ($\phi_+$ and $\phi_-$ respectively) and moving to their symmetric and anti-symmetric combinations (defined by $\phi_\pm=\phi\pm\hat\phi$ and often termed ``classical'' and ``quantum'' components), we have:
\begin{align}
&2\int_{\gamma_K} dt~\cos(\sqrt{2}\phi)=2\int dt~\cos(\sqrt{2}\phi_+)-\cos(\sqrt{2}\phi_-)\\\nn
&=4\int dt~\sin(\sqrt{2}\phi)\sin(\sqrt{2}\hat\phi)=\sum_{\e=\pm,~\hat\e=\pm}\int dt~e^{\sqrt{2}i\e\phi+\sqrt{2}i\hat\e\hat\phi}\;.
\end{align}
Expanding $e^{i S}$ in a Taylor series around $j_\perp=0$ we obtain
\begin{align}
&C_{12}(t)  \equiv \av{\cos(\sqrt{2}\phi(t)~)}\nn\\
&={\rm Re} ~\av{e^{\sqrt{2}i\phi}\sum_{N}\frac{\left(i~j_\perp\rho_0\right)^N}{N!}\left(\sum_{\e,\hat\e}\int_0^\infty dt\int_{-\infty}^\infty dx~e^{\sqrt{2} i\left(\e \phi+\hat\e\hat\phi\right)}\right)^N}\label{eq:eiphi1}\;.
\end{align}
Note that the $N$-th term of Eq.~(\ref{eq:eiphi1}) contains $2N$ continuous integrals and $2N$ discrete sums, whose dummy indexes we will denote as  $(x_i,~t_i,~\e_i,~\hat \e_i)$ with $i=1,...,N$. In analogy to the equilibrium case \cite{frohlich81,gritsev06}, we can interpret the first three elements as the space-time coordinates and charges of a Coulomb gas with $N$ particles (dual to the atoms of the original problem). In contrast, the $\hat\e_i$ indexes do not have an equilibrium analogue and rather impose a light-cone constraint \citeSM, limiting all contributing ``charges'' to the past-light-cone of $(x,t)$, or $|x - x_i| < t-t_i$. This 
light-cone effect is generic to global quantum quenches (see for example Refs.~\cite{calabrese06,mathey10}) and prevents the exact mapping to an equilibrium model.

\begin{figure}[t]
\includegraphics[scale=0.8]{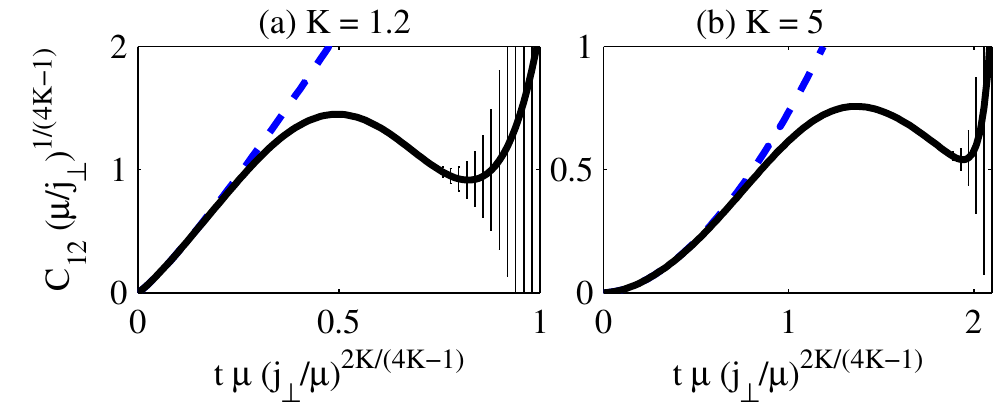}%\vspace{-0.3cm}
\caption{\coloronline Series expansion (\ref{eq:series}). Numerical results obtained by keeping the first $N_{\rm max}$ terms of the series (solid curve and error-bars) and analytic results \citeSM~for the first-order contribution $N=1$ (dashed curve).  (a) $K=1.2,~N_{\rm max}=11$; (b) $K=5,~N_{\rm max}=23$. In both cases we used $10^6$ random evaluation points of $\{(x_i,~t_i)\}_{i=1}^N$ .}\label{fig:montecarlo}%\vspace{-0.3cm}

\end{figure}

The expectation value appearing in Eq.~(\ref{eq:eiphi1}) refers to the initial state, corresponding to the ground state of two independent Luttinger liquids, and can be computed using the quadratic part of Eq.~(\ref{eq:SSG}). After rescaling the time and space variables of integration \citeSM, we arrive to
\begin{align}
C_{12}(t) &= \sum_{N} c_N \left(\frac{j_\perp}{\mu}\right)^{N} \left(\mu t\right)^{2N - (N+1)/(2K)}\\
&= \left(\frac{j_\perp}{\mu}\right)^{1/(4K-1)} \sum_{N} c_N\left[(j_\perp/\mu)^{2K/(4K-1)} \mu t\right]^{2N - (N+1)/(2K)}\label{eq:series}\;.
\end{align}
The coefficients $c_N$ are universal: they do not depend neither on the cutoff nor on $j_\perp$.

To achieve a universal scaling of the form of Eq.~(\ref{eq:scalinglaw}) it is necessary to rescale the time in units of the inverse gap. For the Sine-Gordon model, the excitation gap can be exactly computed using the form-factor approach \cite{zamoldochikov95} and, in our notations, it is proportional to $\D = \mu(j_\perp/\mu)^{2K/(4K-1)}$, leading to
\begin{align}
C_{12}(t) &= \left(\frac{\D}{\mu}\right)^{\eta} \sum_{N} c_N~\Big(\,\D t\,\Big)^{2N - (N+1)/(2K)}\equiv \left(\frac{\D}{\mu}\right)^{\eta} R\Big(\D t\Big)\;. \label{eq:scalinglaw2}
\end{align}
with $\eta=1/(2K)$. Remarkably, we obtain the same ``anomalous'' scaling dimension $\eta$ as in the ground state of the model \cite{lukyanov97}. This result corroborates with a scaling analysis \cite{degrandi10} showing that the average amount of energy-per-mode introduced by the quench scales to zero much faster than the gap. 
%Also note that as expected, in the limit of non-interacting bosons $K\to\infty$, we restore the mean field result presented above, $\eta=0$.

%As anticipated, the dynamics is dominated by quantum effects. This is of course not sufficient to prove a complete thermalization of the steady state \cite{rigol08}, for which comparison with higher order correlations is needed.

In the above derivation we used the asymptotic values of correlation and response functions, related to the low-frequency modes, and we neglected the contributions from the high-frequency modes. These non-universal contributions are analytic functions of the bare parameters. As such, they can at most contribute to $C_{12}$ a term that is linearly proportional to $j_\perp$. In contrast, the universal contributions scale at long times as $j_\perp^{1/(4K-1)}$ and are therefore dominant for any $K>1/2$. Remarkably, the intermediate case of $K=1/2$ can be exactly solved by mapping the problem to non-interacting fermions  \cite{iucci10}. Indeed, at this point, the universal and non-universal contributions scale in the same way and the latter are no more negligible.

At present, we were not able to compute the universal function $R$ by the analytic resummation of the series (\ref{eq:series}). Instead, we numerically estimate its first few terms by  averaging over random samplings of $(\e_i,~x_i,~t_i)$. The resulting universal curves are shown in Fig.~\ref{fig:montecarlo} for two different values of $K$. Our calculations are presently limited to time scales of order $1/\Delta$ due to a ``dynamical sign problem'', related to the alternating signs of the coefficients $c_N$. To predict the universal behavior at longer times, we will now present two alternative numerical methods.

\begin{figure}[t!]
\includegraphics[scale=0.8]{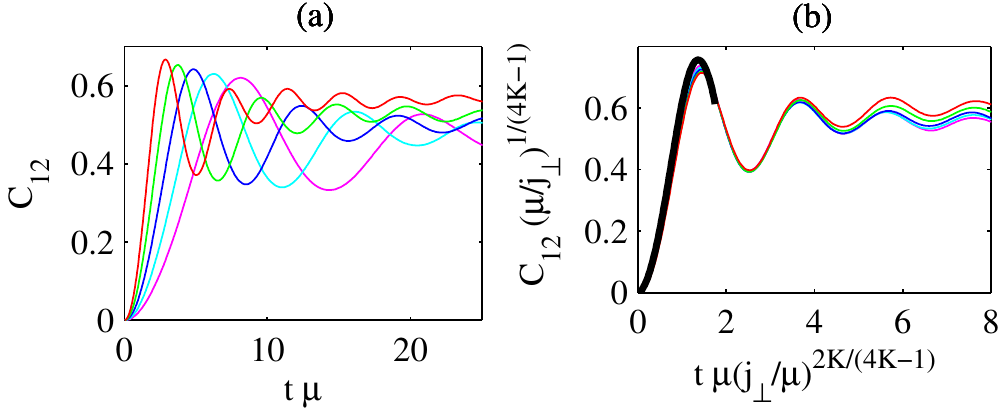}%\vspace{-0.2cm}
\caption{\coloronline 
%Time evolution of the interference contrast $C_{12}=\cos(\sqrt{2}\phi)$ obtained from the 
Truncated Wigner approach (\ref{eq:semiclassic}). (a) Numerical results for $j_\perp/\mu$= 0.27 {\rm (red)},~0.16 {\rm (green)},~0.10 {\rm (blue)},~0.060 {\rm (cyan)},~0.036 {\rm (magenta)} with $K=5$, system size $L=400$, number of random paths $M=400$. (b) Data-collapse according to Eq.~(\ref{eq:scalinglaw2}), superimposed to the result from the series expansion (black curve, identical to Fig.~\ref{fig:montecarlo}(b)).} \label{fig:semiclassic}
\end{figure}

The first method extends the semiclassical approach presented above for the two-mode model. The corresponding classical equation of motion is \cite{clemens12} 
%(obtained by variating the action with respect to the quantum component $\hat\phi$). In this case, the equation of motion are simply
\be \partial^2_t \phi =\partial_x^2 \phi - \frac{2\pi}{K} j_\perp\rho_0\sin(\sqrt{2}\phi~)\;. \label{eq:semiclassic}\ee
The numerical solution of this equation is in good agreement with the series expansion and extends to longer times, as shown in Fig.~\ref{fig:semiclassic} for $K=5$. Note however that Eq.~(\ref{eq:semiclassic}) is characterized by a single energy scale $\sqrt{2\pi j_\perp\rho_0/K}\sim j_\perp^{1/2}$, matching the excitation gap of the original quantum model only for $K\gg 1$. As expected, the semiclassical approach breaks down for $K\sim1$, where the correspondent equilibrium system approaches a quantum phase transition.

\begin{figure}[t!]
\includegraphics[scale=0.8]{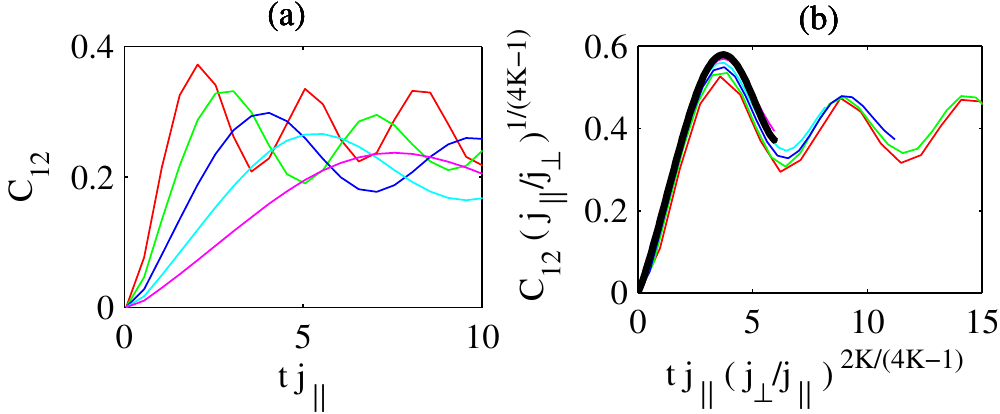}
\caption{\coloronline 
%Time evolution of the interference contrast $C_{12}=\av{\cos(\sqrt{2}\phi)}$ obtained from the 
Hard-core bosons model (\ref{eq:Htebd}). (a) Numerical results for $j_\perp/j_\parallel$= 0.27 {\rm (red)},~0.16 {\rm (green)},0.10 {\rm (blue)},~0.060 {\rm (red)},~0.036 {\rm (cyan)},~0.022 {\rm (magenta)}, with $V/j_\parallel=-1/2$, or $K = \pi/(2\pi-2\arccos(V/2j_\parallel )) \approx 1.2$. The calculations were performed using the Time Evolving Block Decimation (TEBD) algorithm of Refs.~\cite{vidal93,openTEBD} for system size $L=100$, number of states $\chi=100$ and took around three days/curve on a single core. (b) Data collapse  according to Eq.~(\ref{eq:scalinglaw2}), superimposed to the result from the series expansion (black curve, identical to Fig.~\ref{fig:montecarlo}(a) after the appropriate rescaling)%to account for the mapping from the microscopic model (\ref{eq:Htebd}) to the effective field theory (\ref{eq:HSG}))
.}\label{fig:TEBD}
\end{figure}

%{\it Lattice model--} 
To describe the regime $K\sim1$, we map the problem into a lattice model and apply one of the derivatives of the Density Matrix Renormalization Group (DMRG) algorithm \cite{white92,scholl06}. This approach is well suited to situations, as the one under present consideration, in which the time evolution is performed with a {\it gapped} Hamiltonian. To minimize the memory requirements we consider hard-core bosons on a ladder (see Fig.~\ref{fig:schematic}(c)) represented by the Hamiltonian 
\begin{align} H(t)=\sum_{L=1}^N & \left[\sum_{\a=1,2} - j_\parallel \left(b_{\a,i}\yd b_{\a,i+1} + {\rm H.c.}\right) + V n_{\a,i}n_{\a,i+1}\right]\nn \\&- \Theta(t) j_{\perp}\left(b_{1,i}\yd b_{2,i} + {\rm H.c.}\right)\;.\label{eq:Htebd}\end{align}
Here the energy scale $j_\parallel$ sets the chemical potential and $V<0$ is a nearest neighbor {\it attractive} interaction, meant to partially counter-balance the hard-core constraint \footnote{In actual experiments with soft-core bosons such term is not necessary}. For $j_\perp=0$ the two independent chains are Bethe-ansatz \cite{bethe31} solvable, allowing to exactly determine the Luttinger parameter \cite{haldane80,sirker06}. The numerical simulation of the model (\ref{eq:Htebd}) is presented in Fig.~\ref{fig:TEBD} and confirms our scaling ansatz (see the caption for details).

%v&=&\frac{\pi\sqrt{1-(V/2J)}}{2\arccos(V/2J)}\\
%
%Relation between K and K_{sirker}=\frac{\pi}{\pi-\arccos(V/2J)}
%K_{\phi,\rm single}&=&K_{sirker}/2   --------- (The AFM transition point is at K_{\phi,\rm single}=1/2$
%K_{\phi,\rm single}=1/(4K_{\phi,\rm single}  --------- (2K becomes 1/2K) 
%K_\phi =2 K_{\phi,\rm single}  ------- (The correlation function exponent is twice large)
%==> K_\phi= \frac{1}{K_{\phi,\phi,\rm single}}
%==> K_NEW = K_{\phi,\phi,\rm single}/2
%Also Check: at the AFM transition $K_\phi=1/2$, meaning that ${e^{i\phi_1}e^{-i\phi_1}} ~ x^{-1/(2K)}=1/x and $C_12(x,t) ~ 1/x^2$, or $K=1$.

%{\it semiclassical approach --} 

{\em Discussion and conclusion --} In contrast to the above-mentioned models, actual experiments necessarily involve a finite initial temperature $T_0$ and a finite quench time $\tau$. 
%In actual experiments, one necessarily deals with finite initial temperature $T_0$ and a finite ramping time $\tau$. 
Intuitively, one may expect the idealized model of a sudden quench to hold only as long as $\tau$ is smaller than the inverse of the energy of the highest excited state. For the Hamiltonian (\ref{eq:HMF2}) this energy scale is $N_{\rm excited} \D \sim N j_\perp$, leading to the requirement $\tau\ll1/(Nj_\perp)$. However, as discussed in Ref.~\cite{witthaut06}, the most probable excitation path does not directly connect the ground state to the highest excited state, but is rather given by subsequent steps connecting neighboring levels, each with energy splitting $\Delta$. This observation indicates that the  more lenient condition $\tau\ll1/\D$  is sufficient to observe a universal behavior. A similar scaling argument holds for the initial temperature $T_0$, which needs to be smaller than $\D$. These two constraints, $\tau\ll1/\D$ and $T\ll\D$, should be combined with the weak coupling condition $\sqrt{j_\perp/\mu}=\D/\mu\ll1$, requiring temperatures (ramp times) smaller than the (inverse) chemical potential by one order of magnitude. These conditions are in the reach of present experiments with ultracold neutral atoms (see also  the Appendix for the details of a possible realization with wave-interferometers of atoms on a chip \cite{schumm05,gring12}).

A final and important question regards the coupling to the modes that have been neglected in the model Hamiltonians (\ref{eq:HMF}) and (\ref{eq:HSG}). In particular, the former does not include Bogoliubov excitations and the latter neglects the coupling to the ``symmetric'' mode $\bar\phi = (\phi_1+\phi_2)/\sqrt{2}$. These low-frequency modes may act as a dissipative bath for the relative-phase $\phi$ and lead to a further ``rephasing''. A complete analysis of their effect goes beyond the scope of the present Letter. At this stage we can only comment that, based on the analogy with the equilibrium case, one should expect the couplings to these additional modes to be irrelevant in an RG sense and thus to effect the dynamics only at time scales much larger than $\Delta$. Accordingly, the life-time of the predicted ``prethermalized'' state should tend to infinity as one approaches the weak-coupling limit $j_\perp/\mu\to0$.

%{\it Summary --} 
In summary, we studied the many-body quantum dynamics of two initially independent condensates, suddenly coupled via uniform tunneling (``global quench''). In two and three dimensions we applied a ``two-mode'' model, while in one dimension we mapped the problem to a gapped Sine-Gordon model. Both theories are strongly interacting and lead to non-perturbative dynamics. We determined the scaling laws for expectation values of physical observables and predicted a universal time-dependence, connecting short and long times. Our findings may have important implications for both numerical calculations and experiments, in which the long-time evolution is hard to achieve. They may also allow to perform finite-size scaling and renormalization-group analysis in the time domain \cite{mitra11, dallatorre11, mitra12, mitra12b}, in analogy to the established equilibrium techniques.

%\newpage

We wish to thank J. Simon for many useful insights on the experimental realization and C. De Grandi, D. Huse, C. Karrasch, D. Schuricht, Shuyuan Wu for critically reading the manuscript. We also acknowledge useful discussions with  E. Altman, E. Berg, I. Bloch, T. Giamarchi, P. Le Doussal, D. Orgad, C. Neuenhahn. We acknowledge support from Harvard-MIT CUA, NSF Grants No. DMR-07-05472 and DMR-09-07039, DARPA OLE program, AFOSR Quantum Simulation MURI and AFSOR FA9550-10-1-0110, the ARO-MURI on Atomtronics, Sloan Foundation, Simons Foundation. Finally, we would like to thank the KITP for hospitality during the workshop on "Quantum Dynamics in Far from Equilibrium Thermally Isolated Systems"%
.

%\newpage

%\bibliographystyle{unsrt}

\begin{appendix}
\section{Derivation of the semiclassical equations of motion from Eq.~(3)}%\ref{eq:HMF2})}
Consider the generic quantum time-dependent Hamiltonian \be H(t) = \frac{\mu}{2N} p^2 + 2 N j_\perp f(t) V(x),\ee 
where $x,~p$ are unitless canonical variables, and $V(x),~f(t)$ are unitless functions. For the ramp of a Josephson junction , $V(x) = \cos(x)$ and 
\be f(t) = \left\{\ba{c}t/\tau\quad t<\tau\\ 1 \quad t>\tau\ea\right. \ee The correspondent action is:
\be S[x,p] = S_0 + i \int_{\gamma_K}dt~ p\partial_t x - \frac{\mu}{2N} p^2 - 2 N j_\perp f(t) V(x) ,\ee
where $\gamma_K$ runs over the contour $(0\to +\infty \to 0)$ and $e^{iS_0} = \braket{\psi_0}{p^+_0,x^+_0}\braket{p^+_0,x^+_0}{\psi_0}$ sets the initial conditions. If the initial state is $\ket{p=0}$, we have \be e^{iS_0} = \delta(p^+_0)\delta(p^-_0)\delta(x^+-x^-)\ee
After integrating-out the $p$ variable we get:
\be S[x,p] = S_0 + i N\int_{\gamma_K}dt \frac{1}{2\mu} (\partial_t x)^2 - 2 j_\perp f(t) V(x) \ee
In the limit of $N\to \infty$, only the saddle point contributes to the path integral. Here, the saddle point corresponds to the classical equation of motions
\be \partial_t^2 x = 2\w_0 f(t) V'(x) \ee
with initial conditions substituted by the distribution function $P(x_0,t_0) = e^{iS_0[x_0,\partial_t x_0]}$. Note that the microscopic energy scales $J = j_\perp N$ and $U = \mu/N$ completely disappeared from the problem.

\section{Technical details of the ``Coulomb gas'' expansion Eq.~(8)}%(\ref{eq:series})} 

The starting point of the present derivation is  Eq.~(7)%~(\ref{eq:eiphi1})
. If we define $\phi_0=\phi(x,t)$, $\e_0=\hat\e_0=1$ and use the compact notation $\sum_N=\sum_{N,\{\e_i,\hat\e_i\}}$, and $\int_N = \prod_{i=1}^N \int_0^\infty dt_i \int_{-\infty}^\infty dx_i $, we can express the interference contrast as
\begin{align}
C_{12}(t)
=&\sum_N \frac{(i j_\perp\rho_0)^N}{N!}\int_N \e_i \hat\e_i~\av{\exp\left(\sqrt{2}i\sum_{i=0}^N \e_i\phi_i + \hat\e_i\hat\phi_i\right)}\label{eq:eiphi2}
\end{align}

The expectation value refers to the initial state, given by the ground state of two independent Luttinger liquids, and can be computed from the action (5) %(\ref{eq:SSG}) 
with $j_\perp=0$. Dealing with a quadratic action, we can apply Wick's theorem
$\av{e^{iA}}=\av{e^{-\av{A^2}/2}}$ and obtain
\begin{align}
C_{12}(t)
=\sum_N &\frac{(i j_\perp\rho_0)^N}{N!}\int_N  \e_i \hat\e_i~\exp\left(-\av{~\left(\sum_{i=0}^N \e_i\phi_i + \hat\e_i\hat\phi_i\right)^2}\right)\\
=\sum_N &\frac{\left( i j_\perp \rho_0\right)^N}{N!}\int_N \e_i~\exp\left(-\sum_{i,j=0}^N \e_i\e_j\av{\phi_i\phi_j}\right)\nn\\
&\times~\hat \e_i~\exp\left(-\sum_{i,j=0}^N \e_i\hat\e_j\av{\phi_i\hat\phi_j}\right),\label{eq:eiphi5}
\end{align}
where we used the identity $\av{\hat\phi~\hat\phi}=0$, related to the causality constraint \cite{kamenev_book}.

In one dimension, the correlation functions $\av{\phi_i \phi_j}$ are infra-red divergent. As a consequence the above expression vanishes, unless $\sum_{i=0}^N \e_i = 0$. If we interpret $\e_i$ as the charges of a Coulomb gas, this condition coincides with the requirement of a total charge neutrality. It constraints $N+1$ to be even and $\prod_i \e_i = (-1)^{(N+1)/2}$. This alternating sign will be the origin of a ``sign problem'', preventing the numerical resummation of the series.

We now assume that the largest contribution to integrals over time and space comes from the universal asymptotic behavior of the correlation functions \cite{luther74}
\begin{align} 
\av{(\phi_i-\phi_j)^2}&=\frac{1}{8K}\log\left(\mu\left|(x_i-x_j)^2-(t_i-t_j)^2\right|\right)\nn\\
\av{\phi_i\hat\phi_j}&=\frac{\pi}{4K} \Theta\left(t_i-t_j-\left|x_i-x_j\right|)\right)\label{eq:phiphi}
\end{align} 
We discussed the non-universal short-distance behavior at the end of the paper and show that they indeed are negligible for $K>1/2$. 

Using Eq.~(\ref{eq:phiphi}), we obtain that the second line of Eq.~(\ref{eq:eiphi5}) becomes
\be \hat \e_j \exp\left(-\hat\e_j\sum_{i=0}^N \e_i\av{\phi_i\hat\phi_j}\right) = -2 i \sin\left(n_j \frac{\pi}{4K}\right).\ee
Here we explicitly performed the sum over $\hat\e_j$ and we defined $n_j$ as the total charge of the points inside the future light cone of $(x_j,t_j)$. Note that $C_{12}(t)$ vanishes unless $n_j\neq0$ for all $j=1,..,N$. This condition can be fulfilled only if all the ``charges'' are inside the past light-cone of $(x,t)$, satisfying $|x-x_j|<t-t_j$. This ``light-cone physics'' is relevant to any global quantum quench (see f.e. Refs.~\cite{calabrese06,mathey10}) and highlights the difference between similar equilibrium Coulomb gas expansions.

The first line of Eq.~(\ref{eq:eiphi5}) equals to
\be \exp\left(-2\sum_{i,j=0}^N \e_i\e_j\av{\phi_i\phi_j}\right) = \prod^N_{i,j=0,i\neq j}\left|(x_i - x_j)^2 - (t_i - t_j)^2\right|^{\e_i\e_j/(4K)}, \ee
Note that there are $(N+1)N$ homo-sign combinations and $(N+1)(N+1)$ hetero-sign combinations, thus at long times this term is proportional to $\left(t^{1/2K}\right)^{(N+1)N/2-(N+1)(N+1)/2} = t^{-(N+1)/(2K)}$. To obtain Eq.~(8) %(\ref{eq:series}) 
it is sufficient to rescale the time and space integration parameters $dt \to t~dt$ and $dx \to t~dx$.

The coefficients $c_N$ are universal (do not depend neither on the cutoff nor on the gap) and are given by 
\begin{align} c_N = &\sum_{\{\e_i=\pm\}} \frac{2^N (-1)^{\frac{N+1}2}}{N!} \prod_{i=1}^N \int_{-\infty}^\infty dx_i \int_0^\infty dt_i\nn  \\&
\sin(-n_i \frac{\pi}{2 K})\prod^N_{j \neq i}\left|(x_i - x_j)^2 - (t_i - t_j)^2\right|^{\e_i\e_j/(4K)}.\end{align}
with $(x_0,t_0)=(0,1)$ and $\epsilon_0=1$. The coefficient $c_1$ corresponds to the first order time-dependent perturbation theory and can be computed analytically
\begin{align}
c_1 &= 2\sin\left(\frac{\pi}{2 K}\right) \int_{0}^\infty dt \int_{-t}^t dx~\frac{1}{|x^2-t^2|^{1/(2K)}}\\
&= \frac{2\sin\left(\frac{\pi}{2K}\right)~\sqrt{\pi}~\Gamma\left(1-\frac{1}{2K}\right)}{\left(2-\frac1K\right)~\Gamma\left(\frac32-\frac1{2K}\right)},\label{eq:c1} 
\end{align}
Here the second line has been computed with Wolfram's Mathematica (v8.0.1) and $\Gamma$ is the Gamma function.

\begin{figure}[t]
\centering
\includegraphics[scale=0.8]{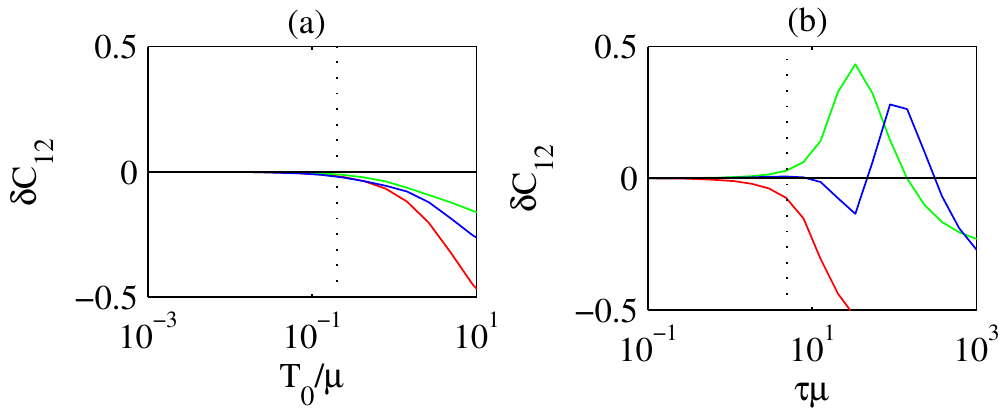}
\caption{\coloronline Comparison between the idealized case considered in Fig.~2 %\ref{fig:meanfield}
 and (a) finite initial temperature $T_0$, or (b) finite ramp time $\tau$. The solid lines represent $\delta C_{12}(t) \equiv C^{\rm finite}_{12}(t)-C^{\rm ideal}_{12}(t)$ at different times $t=10/\mu$ (red), $20/\mu$ (green), and $30/\mu$ (blue). The vertical dashed lines correspond to (a) $T_0=\D$, (b) $\tau=1/\D$. As predicted, the dynamics on the left side of these lines is independent of $T_0$ and $\tau$. {\it Numerical parameters --} $j_\perp/\mu=0.01$, ~ $\D=2\sqrt{j_\perp/\mu}=0.2\mu$, and $N=1000$.}
\label{fig:finite}
\end{figure}

\section{Finite temperature and finite ramp time} We now consider the effects of a finite initial temperature $T_0$ and a finite ramping time $\tau$. Because the dynamics described in this paper is characterized by a single time scale $\D^{-1}$ it is natural to assume that the universal dynamics should not depend on these parameters as long as $T_0\ll\D$ and $\tau\ll1/\D$. For the mean-field model (3) %(\ref{eq:HMF}) 
this claim can be verified by exact diagonalization, as shown in Fig.~\ref{fig:finite}, or by the semiclassical method introduced above.

For the one dimensional case, one can move one step further and show that a finite $T_0$ and $\tau$ can be used as additional scaling parameters:
\be
C_{12}(t) =\av{\cos(\sqrt{2}\phi(t))}=\left(\frac{\D}{\mu}\right)^\eta R\left(\D t,~\frac{T_0}{\D},~ \Delta \tau\right)\label{eq:generalized}
\ee
In particular, this scaling relation confirms that, as long as $T_0\ll\D$ and $\tau\ll1/\Delta$, the resulting dynamics should be independent on these parameters. To demonstrate the validity of Eq.~(\ref{eq:generalized}) and starting from the series expansion presented above, one obtains:
\begin{align}
C_{12}(t)
=&\sum_N \frac{(i j_\perp\rho_0)^N}{2^N N!}\int_N \e_i \hat\e_i f\left(\frac{t_i}{\tau}\right)~\av{\exp\left(2i\sum_{i=0}^N \e_i\phi_i + \hat\e_i\hat\phi_i\right)}_{T_0}\label{eq:eiphi3}
\end{align}
Here $f$ determines the shape of the ramp and satisfies $f(0)=0$ and $f(x>1)=1$, and the expectation value $\av{...}_{T_0}$ refers to an equilibrium system at temperature $T_0$. If we now rescale the time and space coordinates in units of $\D^{-1}$ we obtain:
\begin{align}
\sum_N \frac{(i j_\perp\rho_0)^N\D^{2N}}{2^N N!}\int_N \e_i \hat\e_i f\left(\frac{t_i}{\D\tau}\right)~\av{\exp\left(2i\sum_{i=0}^N \e_i\phi'_i + \hat\e_i\hat\phi'_i\right)}_{T_0},\label{eq:eiphi4}
\end{align}
where we define $\phi'_i=\phi(\D x_i,\D t_i)$. Due to the scale invariance of the underlying theory (and provided that $T_0\ll\mu$) we have that
\be \av{\phi'_i\phi'_j}_{T_0} \approx \av{\phi_i\phi_j}_{T_0/\D} + \frac1{8K}\log\left(\frac{\D}{\mu}\right)\label{eq:invariance}\ee
Combining this result with Eq.~(\ref{eq:eiphi4}) one can easily show the validity of the generalized scaling relation (\ref{eq:generalized}). This scaling relation is expected to be valid in the mean-field model as well, corresponding to the $K\to\infty$ limit of the one-dimensional case.

%\vspace{0.5cm}

\begin{figure*}[t!]
\centering
\includegraphics[scale=0.8]{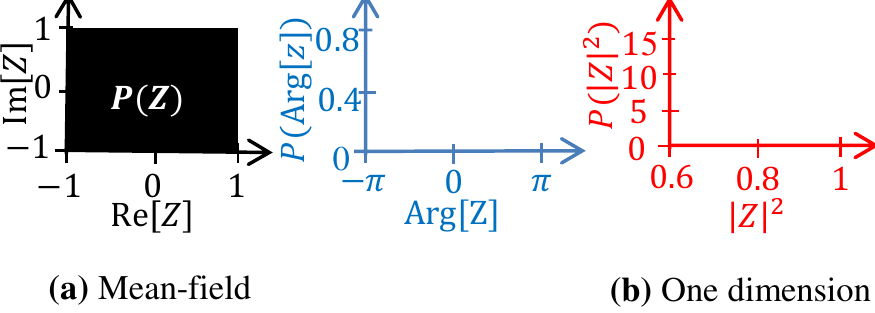}
\begin{tabular}{c c}
\includegraphics[scale=0.8]{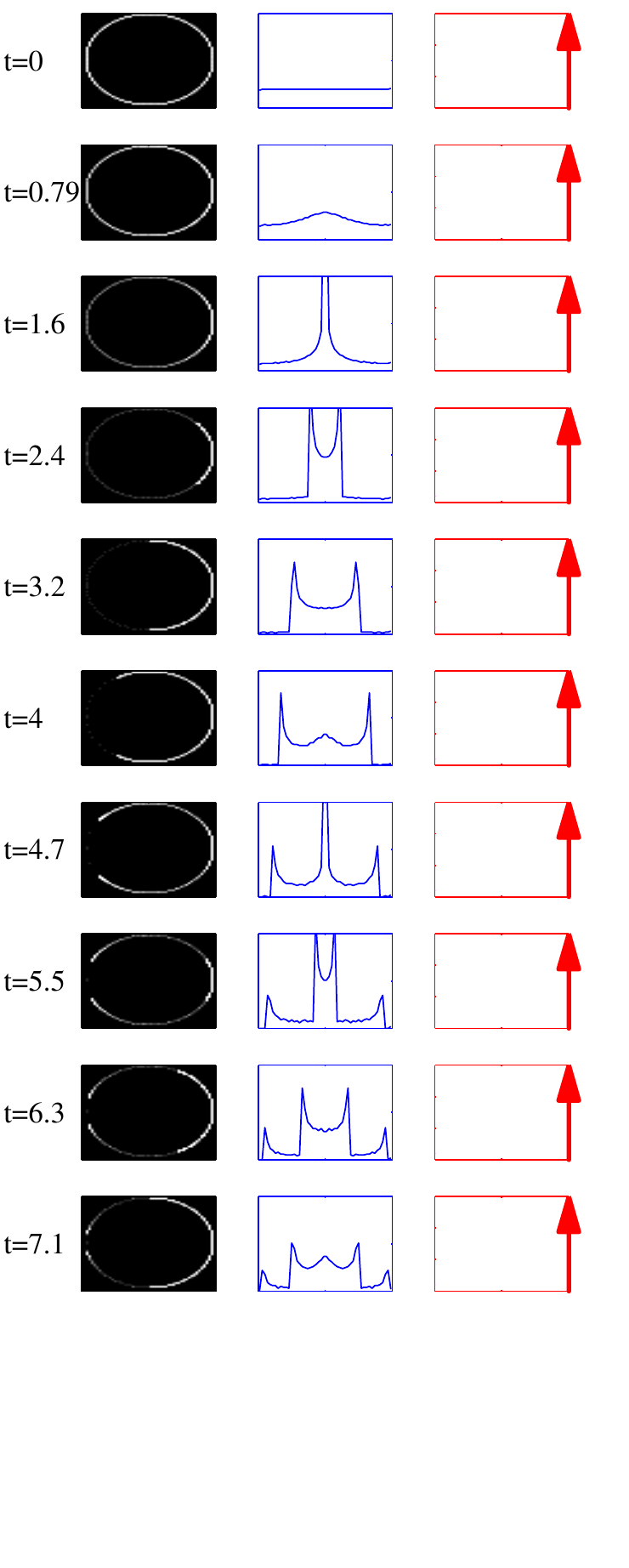}&\includegraphics[scale=0.8]{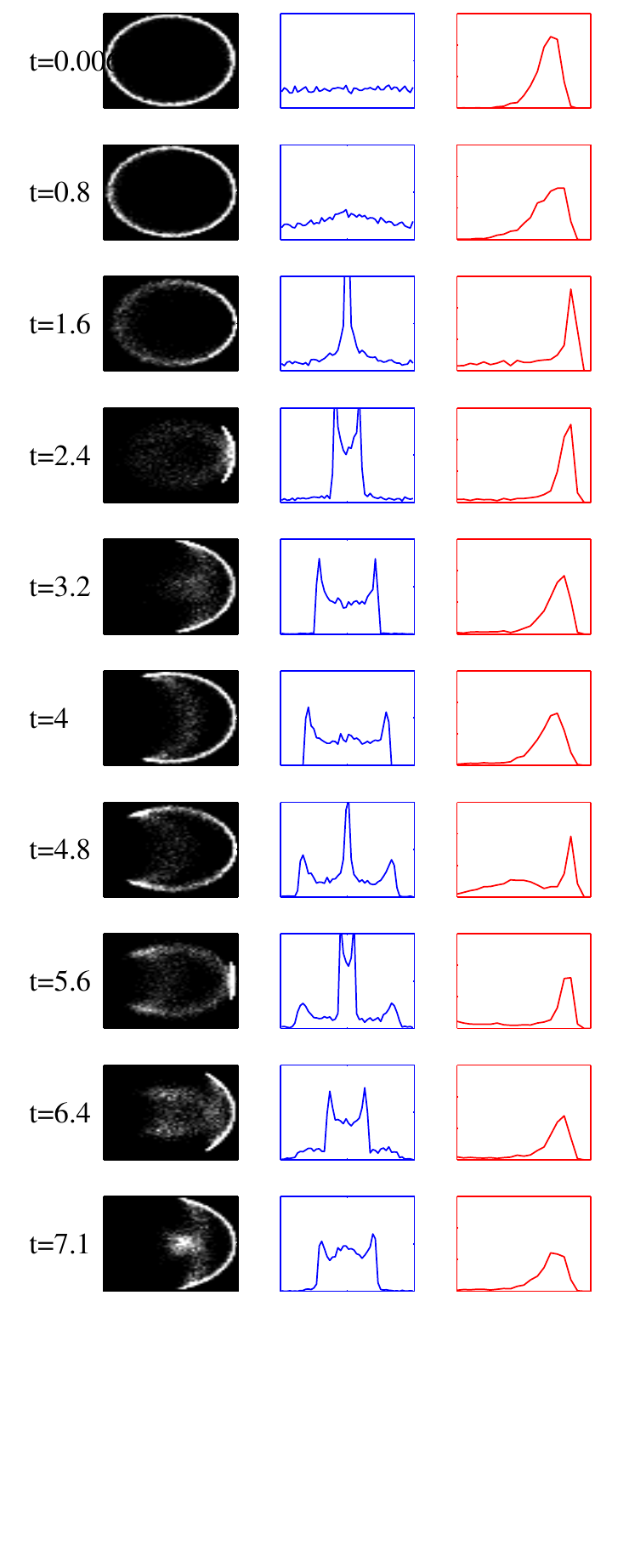}
\end{tabular}
\vspace{-2.5cm}
\caption{\coloronline Full distribution function of the spatially integrated coherence (\ref{eq:Z})
, according  to the semiclassical equations of motion of (a) the two-mode model and (b) the one dimensional model (11) %(\ref{eq:semiclassic})
. Both subplots represent in the first column a colorplot of the distribution function of $Z$ on the imaginary plane. In the second column, the distribution function of the phase of $Z$. In the third column, the distribution function of the absolute value square of $Z$. {\it Numerical values --}  (a) $j_\perp/\mu=0.1$, time is given in units of the inverse plasma frequency $\D=2(j_\perp \mu)^{-1/2}$. (b) $j_\perp/\mu=0.1,~K=25,~L=400$, corresponding to the scaling limit $j_\perp/\mu\ll1$,~$K\gg1$,~$\D L \gg 1$. Time is given in units of the inverse gap $\D^{-1}\approx \Big((4\pi/K) j_\perp \rho_0 \Big)^{-1/2}$.}
\label{fig:full}
\end{figure*}

\section{Experimental realization} 

A natural realization of the proposed experiment involves wave-interferometers of atoms on a chip \cite{schumm05}. This set-up has been successfully used to realize ``dephasing'' experiments, in which one condensate is suddenly split in two parts \cite{schumm05,gring12}. There, the time evolution is described by a gapless Hamiltonian and the resulting dynamics can be studied by the exact solution of the quadratic Luttinger liquid Hamiltonian \cite{gritsev06,bistritzer07,gring12}. Due to the absence of an excitation gap, at long-times the ``dephasing'' process is primarily thermal, rather than quantum \cite{barmettler09, gring12}.  In contrast, in the present ``rephasing'' experiment we predict a universal dynamics dominated by quantum effects both at short and long times. %with respect to the inverse gap $\Delta$.

This type of interferometry poses a serious challenge for the observation of one dimensional physics, due to the extremely high value of the Luttinger parameter $K\approx 50$. For this system $\eta = 1/(2K) \approx0$ and the deviations of the average interference $C_{12}(t)$ from predictions of the two-mode model (\ref{eq:HMF}) are extremely small and hard to observe. %(compare for example Fig.~\ref{fig:meanfield} with Fig.~\ref{fig:semiclassic}).
As suggested in Ref.~\cite{gritsev06}, the one dimensional nature of the problem can be observed by probing the spatially integrated coherence
\be Z = \frac{1}{L}\int_{-L/2}^{L/2} dx~e^{i\sqrt{2}\phi(x,t)}\,,\label{eq:Z}\ee
where $L$ is the system size. In the two subplots of Fig.~\ref{fig:full} we compare the evolution of the full-distribution function of $Z$ for respectively the two-mode and the one-dimensional models. In each subplot, the first column contains a colorplot, proportional to the probability of observing a specific complex value of the operator $Z$. The second column refers to the phase of $Z$. The two-mode and one-dimensional cases look identical for short times, but differ at longer times. In particular, the side-peaks at $\phi\approx\pm\pi$ disappear as a consequence of the one dimensional dynamics. The third column represents the probability distribution of the absolute value of $Z$. For the two-mode case, it simply corresponds to a delta-function peak at the maximal value of $|Z|=1$. In contrast, for the one-dimension case $P(|Z|)$ oscillates between the initial Gumbel distribution and narrower peaks, indicating the formation of long-range order in the relative phase.

\end{appendix}

\bibliographystyle{aipnum4-1}

\bibliography{SGquench}

\begin{thebibliography}{48}%
\makeatletter
\providecommand \@ifxundefined [1]{%
 \@ifx{#1\undefined}
}%
\providecommand \@ifnum [1]{%
 \ifnum #1\expandafter \@firstoftwo
 \else \expandafter \@secondoftwo
 \fi
}%
\providecommand \@ifx [1]{%
 \ifx #1\expandafter \@firstoftwo
 \else \expandafter \@secondoftwo
 \fi
}%
\providecommand \natexlab [1]{#1}%
\providecommand \enquote  [1]{``#1''}%
\providecommand \bibnamefont  [1]{#1}%
\providecommand \bibfnamefont [1]{#1}%
\providecommand \citenamefont [1]{#1}%
\providecommand \href@noop [0]{\@secondoftwo}%
\providecommand \href [0]{\begingroup \@sanitize@url \@href}%
\providecommand \@href[1]{\@@startlink{#1}\@@href}%
\providecommand \@@href[1]{\endgroup#1\@@endlink}%
\providecommand \@sanitize@url [0]{\catcode `\\12\catcode `\$12\catcode
  `\&12\catcode `\#12\catcode `\^12\catcode `\_12\catcode `\%12\relax}%
\providecommand \@@startlink[1]{}%
\providecommand \@@endlink[0]{}%
\providecommand \url  [0]{\begingroup\@sanitize@url \@url }%
\providecommand \@url [1]{\endgroup\@href {#1}{\urlprefix }}%
\providecommand \urlprefix  [0]{URL }%
\providecommand \Eprint [0]{\href }%
\providecommand \doibase [0]{http://dx.doi.org/}%
\providecommand \selectlanguage [0]{\@gobble}%
\providecommand \bibinfo  [0]{\@secondoftwo}%
\providecommand \bibfield  [0]{\@secondoftwo}%
\providecommand \translation [1]{[#1]}%
\providecommand \BibitemOpen [0]{}%
\providecommand \bibitemStop [0]{}%
\providecommand \bibitemNoStop [0]{.\EOS\space}%
\providecommand \EOS [0]{\spacefactor3000\relax}%
\providecommand \BibitemShut  [1]{\csname bibitem#1\endcsname}%
\let\auto@bib@innerbib\@empty
%</preamble>
\bibitem [{\citenamefont {Mitra}\ \emph {et~al.}(2006)\citenamefont {Mitra},
  \citenamefont {Takei}, \citenamefont {Kim},\ and\ \citenamefont
  {Millis}}]{mitra06}%
  \BibitemOpen
  \bibfield  {author} {\bibinfo {author} {\bibfnamefont {A.}~\bibnamefont
  {Mitra}}, \bibinfo {author} {\bibfnamefont {S.}~\bibnamefont {Takei}},
  \bibinfo {author} {\bibfnamefont {Y.~B.}\ \bibnamefont {Kim}}, \ and\
  \bibinfo {author} {\bibfnamefont {A.~J.}\ \bibnamefont {Millis}},\ }\href
  {\doibase 10.1103/PhysRevLett.97.236808} {\bibfield  {journal} {\bibinfo
  {journal} {Phys. Rev. Lett.}\ }\textbf {\bibinfo {volume} {97}},\ \bibinfo
  {pages} {236808} (\bibinfo {year} {2006})}\BibitemShut {NoStop}%
\bibitem [{\citenamefont {Diehl}\ \emph {et~al.}(2008)\citenamefont {Diehl},
  \citenamefont {Micheli}, \citenamefont {Kantian}, \citenamefont {Kraus},
  \citenamefont {Buchler},\ and\ \citenamefont {Zoller}}]{diehl08}%
  \BibitemOpen
  \bibfield  {author} {\bibinfo {author} {\bibfnamefont {S.}~\bibnamefont
  {Diehl}}, \bibinfo {author} {\bibfnamefont {A.}~\bibnamefont {Micheli}},
  \bibinfo {author} {\bibfnamefont {A.}~\bibnamefont {Kantian}}, \bibinfo
  {author} {\bibfnamefont {B.}~\bibnamefont {Kraus}}, \bibinfo {author}
  {\bibfnamefont {H.~P.}\ \bibnamefont {Buchler}}, \ and\ \bibinfo {author}
  {\bibfnamefont {P.}~\bibnamefont {Zoller}},\ }\href
  {http://dx.doi.org/10.1038/nphys1073} {\bibfield  {journal} {\bibinfo
  {journal} {Nature Physics}\ }\textbf {\bibinfo {volume} {4}},\ \bibinfo
  {pages} {878} (\bibinfo {year} {2008})}\BibitemShut {NoStop}%
\bibitem [{\citenamefont {Dalla~Torre}\ \emph {et~al.}(2010)\citenamefont
  {Dalla~Torre}, \citenamefont {Demler}, \citenamefont {Giamarchi},\ and\
  \citenamefont {Altman}}]{dallatorre10}%
  \BibitemOpen
  \bibfield  {author} {\bibinfo {author} {\bibfnamefont {E.~G.}\ \bibnamefont
  {Dalla~Torre}}, \bibinfo {author} {\bibfnamefont {E.}~\bibnamefont {Demler}},
  \bibinfo {author} {\bibfnamefont {T.}~\bibnamefont {Giamarchi}}, \ and\
  \bibinfo {author} {\bibfnamefont {E.}~\bibnamefont {Altman}},\ }\href
  {http://dx.doi.org/10.1038/nphys1754} {\bibfield  {journal} {\bibinfo
  {journal} {Nature Physics}\ }\textbf {\bibinfo {volume} {6}},\ \bibinfo
  {pages} {806} (\bibinfo {year} {2010})}\BibitemShut {NoStop}%
\bibitem [{\citenamefont {Polkovnikov}\ \emph {et~al.}(2011)\citenamefont
  {Polkovnikov}, \citenamefont {Sengupta}, \citenamefont {Silva},\ and\
  \citenamefont {Vengalattore}}]{polkovnikov11}%
  \BibitemOpen
  \bibfield  {author} {\bibinfo {author} {\bibfnamefont {A.}~\bibnamefont
  {Polkovnikov}}, \bibinfo {author} {\bibfnamefont {K.}~\bibnamefont
  {Sengupta}}, \bibinfo {author} {\bibfnamefont {A.}~\bibnamefont {Silva}}, \
  and\ \bibinfo {author} {\bibfnamefont {M.}~\bibnamefont {Vengalattore}},\
  }\href {\doibase 10.1103/RevModPhys.83.863} {\bibfield  {journal} {\bibinfo
  {journal} {Rev. Mod. Phys.}\ }\textbf {\bibinfo {volume} {83}},\ \bibinfo
  {pages} {863} (\bibinfo {year} {2011})}\BibitemShut {NoStop}%
\bibitem [{\citenamefont {S.~Deng}(2008)}]{deng08}%
  \BibitemOpen
  \bibfield  {author} {\bibinfo {author} {\bibfnamefont {L.~V.}\ \bibnamefont
  {S.~Deng}, \bibfnamefont {G.~Ortiz}},\ }\href@noop {} {\bibfield  {journal}
  {\bibinfo  {journal} {Europhys. Lett.}\ }\textbf {\bibinfo {volume} {84}},\
  \bibinfo {pages} {67008} (\bibinfo {year} {2008})}\BibitemShut {NoStop}%
\bibitem [{\citenamefont {De~Grandi}, \citenamefont {Polkovnikov},\ and\
  \citenamefont {Sandvik}(2011)}]{degrandi11}%
  \BibitemOpen
  \bibfield  {author} {\bibinfo {author} {\bibfnamefont {C.}~\bibnamefont
  {De~Grandi}}, \bibinfo {author} {\bibfnamefont {A.}~\bibnamefont
  {Polkovnikov}}, \ and\ \bibinfo {author} {\bibfnamefont {A.~W.}\ \bibnamefont
  {Sandvik}},\ }\href {\doibase 10.1103/PhysRevB.84.224303} {\bibfield
  {journal} {\bibinfo  {journal} {Phys. Rev. B}\ }\textbf {\bibinfo {volume}
  {84}},\ \bibinfo {pages} {224303} (\bibinfo {year} {2011})}\BibitemShut
  {NoStop}%
\bibitem [{\citenamefont {Chandran}\ \emph {et~al.}(2012)\citenamefont
  {Chandran}, \citenamefont {Erez}, \citenamefont {Gubser},\ and\ \citenamefont
  {Sondhi}}]{chandran12}%
  \BibitemOpen
  \bibfield  {author} {\bibinfo {author} {\bibfnamefont {A.}~\bibnamefont
  {Chandran}}, \bibinfo {author} {\bibfnamefont {A.}~\bibnamefont {Erez}},
  \bibinfo {author} {\bibfnamefont {S.~S.}\ \bibnamefont {Gubser}}, \ and\
  \bibinfo {author} {\bibfnamefont {S.~L.}\ \bibnamefont {Sondhi}},\ }\href
  {\doibase 10.1103/PhysRevB.86.064304} {\bibfield  {journal} {\bibinfo
  {journal} {Phys. Rev. B}\ }\textbf {\bibinfo {volume} {86}},\ \bibinfo
  {pages} {064304} (\bibinfo {year} {2012})}\BibitemShut {NoStop}%
\bibitem [{\citenamefont {Kolodrubetz}, \citenamefont {Clark},\ and\
  \citenamefont {Huse}(2012)}]{kolodrubetz12}%
  \BibitemOpen
  \bibfield  {author} {\bibinfo {author} {\bibfnamefont {M.}~\bibnamefont
  {Kolodrubetz}}, \bibinfo {author} {\bibfnamefont {B.~K.}\ \bibnamefont
  {Clark}}, \ and\ \bibinfo {author} {\bibfnamefont {D.~A.}\ \bibnamefont
  {Huse}},\ }\href {\doibase 10.1103/PhysRevLett.109.015701} {\bibfield
  {journal} {\bibinfo  {journal} {Phys. Rev. Lett.}\ }\textbf {\bibinfo
  {volume} {109}},\ \bibinfo {pages} {015701} (\bibinfo {year}
  {2012})}\BibitemShut {NoStop}%
\bibitem [{\citenamefont {De~Grandi}, \citenamefont {Gritsev},\ and\
  \citenamefont {Polkovnikov}(2010{\natexlab{a}})}]{degrandi10A}%
  \BibitemOpen
  \bibfield  {author} {\bibinfo {author} {\bibfnamefont {C.}~\bibnamefont
  {De~Grandi}}, \bibinfo {author} {\bibfnamefont {V.}~\bibnamefont {Gritsev}},
  \ and\ \bibinfo {author} {\bibfnamefont {A.}~\bibnamefont {Polkovnikov}},\
  }\href {\doibase 10.1103/PhysRevB.81.012303} {\bibfield  {journal} {\bibinfo
  {journal} {Phys. Rev. B}\ }\textbf {\bibinfo {volume} {81}},\ \bibinfo
  {pages} {012303} (\bibinfo {year} {2010}{\natexlab{a}})}\BibitemShut
  {NoStop}%
\bibitem [{\citenamefont {Campos~Venuti}\ and\ \citenamefont
  {Zanardi}(2010)}]{campos10}%
  \BibitemOpen
  \bibfield  {author} {\bibinfo {author} {\bibfnamefont {L.}~\bibnamefont
  {Campos~Venuti}}\ and\ \bibinfo {author} {\bibfnamefont {P.}~\bibnamefont
  {Zanardi}},\ }\href {\doibase 10.1103/PhysRevA.81.032113} {\bibfield
  {journal} {\bibinfo  {journal} {Phys. Rev. A}\ }\textbf {\bibinfo {volume}
  {81}},\ \bibinfo {pages} {032113} (\bibinfo {year} {2010})}\BibitemShut
  {NoStop}%
\bibitem [{\citenamefont {Iyer}\ and\ \citenamefont {Andrei}(2012)}]{yier12}%
  \BibitemOpen
  \bibfield  {author} {\bibinfo {author} {\bibfnamefont {D.}~\bibnamefont
  {Iyer}}\ and\ \bibinfo {author} {\bibfnamefont {N.}~\bibnamefont {Andrei}},\
  }\href {\doibase 10.1103/PhysRevLett.109.115304} {\bibfield  {journal}
  {\bibinfo  {journal} {Phys. Rev. Lett.}\ }\textbf {\bibinfo {volume} {109}},\
  \bibinfo {pages} {115304} (\bibinfo {year} {2012})}\BibitemShut {NoStop}%
\bibitem [{\citenamefont {Karrasch}\ \emph {et~al.}(2012)\citenamefont
  {Karrasch}, \citenamefont {Rentrop}, \citenamefont {Schuricht},\ and\
  \citenamefont {Meden}}]{karrasch12}%
  \BibitemOpen
  \bibfield  {author} {\bibinfo {author} {\bibfnamefont {C.}~\bibnamefont
  {Karrasch}}, \bibinfo {author} {\bibfnamefont {J.}~\bibnamefont {Rentrop}},
  \bibinfo {author} {\bibfnamefont {D.}~\bibnamefont {Schuricht}}, \ and\
  \bibinfo {author} {\bibfnamefont {V.}~\bibnamefont {Meden}},\ }\href
  {\doibase 10.1103/PhysRevLett.109.126406} {\bibfield  {journal} {\bibinfo
  {journal} {Phys. Rev. Lett.}\ }\textbf {\bibinfo {volume} {109}},\ \bibinfo
  {pages} {126406} (\bibinfo {year} {2012})}\BibitemShut {NoStop}%
\bibitem [{\citenamefont {Neuenhahn}, \citenamefont {Polkovnikov},\ and\
  \citenamefont {Marquardt}(2012)}]{clemens12}%
  \BibitemOpen
  \bibfield  {author} {\bibinfo {author} {\bibfnamefont {C.}~\bibnamefont
  {Neuenhahn}}, \bibinfo {author} {\bibfnamefont {A.}~\bibnamefont
  {Polkovnikov}}, \ and\ \bibinfo {author} {\bibfnamefont {F.}~\bibnamefont
  {Marquardt}},\ }\href {\doibase 10.1103/PhysRevLett.109.085304} {\bibfield
  {journal} {\bibinfo  {journal} {Phys. Rev. Lett.}\ }\textbf {\bibinfo
  {volume} {109}},\ \bibinfo {pages} {085304} (\bibinfo {year}
  {2012})}\BibitemShut {NoStop}%
\bibitem [{\citenamefont {Janssen}, \citenamefont {Schaub},\ and\ \citenamefont
  {Schmittmann}(1989)}]{janssen89}%
  \BibitemOpen
  \bibfield  {author} {\bibinfo {author} {\bibfnamefont {H.~K.}\ \bibnamefont
  {Janssen}}, \bibinfo {author} {\bibfnamefont {B.}~\bibnamefont {Schaub}}, \
  and\ \bibinfo {author} {\bibfnamefont {B.}~\bibnamefont {Schmittmann}},\
  }\href {http://dx.doi.org/10.1007/BF01319383} {\bibfield  {journal} {\bibinfo
   {journal} {Zeitschrift fur Physik B Condensed Matter}\ }\textbf {\bibinfo
  {volume} {73}},\ \bibinfo {pages} {539} (\bibinfo {year} {1989})},\ \bibinfo
  {note} {10.1007/BF01319383}\BibitemShut {NoStop}%
\bibitem [{\citenamefont {Zheng}(1998)}]{zheng98}%
  \BibitemOpen
  \bibfield  {author} {\bibinfo {author} {\bibfnamefont {B.}~\bibnamefont
  {Zheng}},\ }\href {\doibase 10.1142/S021797929800288X} {\bibfield  {journal}
  {\bibinfo  {journal} {International Journal of Modern Physics B}\ }\textbf
  {\bibinfo {volume} {12}},\ \bibinfo {pages} {1419} (\bibinfo {year}
  {1998})}\BibitemShut {NoStop}%
\bibitem [{\citenamefont {Kolton}\ \emph {et~al.}(2006)\citenamefont {Kolton},
  \citenamefont {Rosso}, \citenamefont {Albano},\ and\ \citenamefont
  {Giamarchi}}]{kolton06}%
  \BibitemOpen
  \bibfield  {author} {\bibinfo {author} {\bibfnamefont {A.~B.}\ \bibnamefont
  {Kolton}}, \bibinfo {author} {\bibfnamefont {A.}~\bibnamefont {Rosso}},
  \bibinfo {author} {\bibfnamefont {E.~V.}\ \bibnamefont {Albano}}, \ and\
  \bibinfo {author} {\bibfnamefont {T.}~\bibnamefont {Giamarchi}},\ }\href
  {\doibase 10.1103/PhysRevB.74.140201} {\bibfield  {journal} {\bibinfo
  {journal} {Phys. Rev. B}\ }\textbf {\bibinfo {volume} {74}},\ \bibinfo
  {pages} {140201} (\bibinfo {year} {2006})}\BibitemShut {NoStop}%
\bibitem [{\citenamefont {Albano}\ \emph {et~al.}(2011)\citenamefont {Albano},
  \citenamefont {Bab}, \citenamefont {Baglietto}, \citenamefont {Borzi},
  \citenamefont {Grigera}, \citenamefont {Loscar}, \citenamefont {Rodriguez},
  \citenamefont {Puzzo},\ and\ \citenamefont {Saracco}}]{albano11}%
  \BibitemOpen
  \bibfield  {author} {\bibinfo {author} {\bibfnamefont {E.~V.}\ \bibnamefont
  {Albano}}, \bibinfo {author} {\bibfnamefont {M.~A.}\ \bibnamefont {Bab}},
  \bibinfo {author} {\bibfnamefont {G.}~\bibnamefont {Baglietto}}, \bibinfo
  {author} {\bibfnamefont {R.~A.}\ \bibnamefont {Borzi}}, \bibinfo {author}
  {\bibfnamefont {T.~S.}\ \bibnamefont {Grigera}}, \bibinfo {author}
  {\bibfnamefont {E.~S.}\ \bibnamefont {Loscar}}, \bibinfo {author}
  {\bibfnamefont {D.~E.}\ \bibnamefont {Rodriguez}}, \bibinfo {author}
  {\bibfnamefont {M.~L.~R.}\ \bibnamefont {Puzzo}}, \ and\ \bibinfo {author}
  {\bibfnamefont {G.~P.}\ \bibnamefont {Saracco}},\ }\href
  {http://stacks.iop.org/0034-4885/74/i=2/a=026501} {\bibfield  {journal}
  {\bibinfo  {journal} {Reports on Progress in Physics}\ }\textbf {\bibinfo
  {volume} {74}},\ \bibinfo {pages} {026501} (\bibinfo {year}
  {2011})}\BibitemShut {NoStop}%
\bibitem [{\citenamefont {Steel}\ \emph {et~al.}(1998)\citenamefont {Steel},
  \citenamefont {Olsen}, \citenamefont {Plimak}, \citenamefont {Drummond},
  \citenamefont {Tan}, \citenamefont {Collett}, \citenamefont {Walls},\ and\
  \citenamefont {Graham}}]{steel98}%
  \BibitemOpen
  \bibfield  {author} {\bibinfo {author} {\bibfnamefont {M.~J.}\ \bibnamefont
  {Steel}}, \bibinfo {author} {\bibfnamefont {M.~K.}\ \bibnamefont {Olsen}},
  \bibinfo {author} {\bibfnamefont {L.~I.}\ \bibnamefont {Plimak}}, \bibinfo
  {author} {\bibfnamefont {P.~D.}\ \bibnamefont {Drummond}}, \bibinfo {author}
  {\bibfnamefont {S.~M.}\ \bibnamefont {Tan}}, \bibinfo {author} {\bibfnamefont
  {M.~J.}\ \bibnamefont {Collett}}, \bibinfo {author} {\bibfnamefont {D.~F.}\
  \bibnamefont {Walls}}, \ and\ \bibinfo {author} {\bibfnamefont
  {R.}~\bibnamefont {Graham}},\ }\href {\doibase 10.1103/PhysRevA.58.4824}
  {\bibfield  {journal} {\bibinfo  {journal} {Phys. Rev. A}\ }\textbf {\bibinfo
  {volume} {58}},\ \bibinfo {pages} {4824} (\bibinfo {year}
  {1998})}\BibitemShut {NoStop}%
\bibitem [{\citenamefont {Sinatra}, \citenamefont {Lobo},\ and\ \citenamefont
  {Castin}(2001)}]{sinatra01}%
  \BibitemOpen
  \bibfield  {author} {\bibinfo {author} {\bibfnamefont {A.}~\bibnamefont
  {Sinatra}}, \bibinfo {author} {\bibfnamefont {C.}~\bibnamefont {Lobo}}, \
  and\ \bibinfo {author} {\bibfnamefont {Y.}~\bibnamefont {Castin}},\ }\href
  {\doibase 10.1103/PhysRevLett.87.210404} {\bibfield  {journal} {\bibinfo
  {journal} {Phys. Rev. Lett.}\ }\textbf {\bibinfo {volume} {87}},\ \bibinfo
  {pages} {210404} (\bibinfo {year} {2001})}\BibitemShut {NoStop}%
\bibitem [{\citenamefont {Bodet}\ \emph {et~al.}(2010)\citenamefont {Bodet},
  \citenamefont {Est\`eve}, \citenamefont {Oberthaler},\ and\ \citenamefont
  {Gasenzer}}]{bodet10}%
  \BibitemOpen
  \bibfield  {author} {\bibinfo {author} {\bibfnamefont {C.}~\bibnamefont
  {Bodet}}, \bibinfo {author} {\bibfnamefont {J.}~\bibnamefont {Est\`eve}},
  \bibinfo {author} {\bibfnamefont {M.~K.}\ \bibnamefont {Oberthaler}}, \ and\
  \bibinfo {author} {\bibfnamefont {T.}~\bibnamefont {Gasenzer}},\ }\href
  {\doibase 10.1103/PhysRevA.81.063605} {\bibfield  {journal} {\bibinfo
  {journal} {Phys. Rev. A}\ }\textbf {\bibinfo {volume} {81}},\ \bibinfo
  {pages} {063605} (\bibinfo {year} {2010})}\BibitemShut {NoStop}%
\bibitem [{\citenamefont {Giamarchi}(2004)}]{giamarchi_book}%
  \BibitemOpen
  \bibfield  {author} {\bibinfo {author} {\bibfnamefont {T.}~\bibnamefont
  {Giamarchi}},\ }\href@noop {} {\emph {\bibinfo {title} {Quantum Physics in
  One Dimension}}}\ (\bibinfo  {publisher} {Oxford University Press},\ \bibinfo
  {address} {Oxford},\ \bibinfo {year} {2004})\BibitemShut {NoStop}%
\bibitem [{\citenamefont {Haldane}(1980)}]{haldane80}%
  \BibitemOpen
  \bibfield  {author} {\bibinfo {author} {\bibfnamefont {F.~D.~M.}\
  \bibnamefont {Haldane}},\ }\href {\doibase 10.1103/PhysRevLett.45.1358}
  {\bibfield  {journal} {\bibinfo  {journal} {Phys. Rev. Lett.}\ }\textbf
  {\bibinfo {volume} {45}},\ \bibinfo {pages} {1358} (\bibinfo {year}
  {1980})}\BibitemShut {NoStop}%
\bibitem [{\citenamefont {Calabrese}\ and\ \citenamefont
  {Cardy}(2006)}]{calabrese06}%
  \BibitemOpen
  \bibfield  {author} {\bibinfo {author} {\bibfnamefont {P.}~\bibnamefont
  {Calabrese}}\ and\ \bibinfo {author} {\bibfnamefont {J.}~\bibnamefont
  {Cardy}},\ }\href {\doibase 10.1103/PhysRevLett.96.136801} {\bibfield
  {journal} {\bibinfo  {journal} {Phys. Rev. Lett.}\ }\textbf {\bibinfo
  {volume} {96}},\ \bibinfo {pages} {136801} (\bibinfo {year}
  {2006})}\BibitemShut {NoStop}%
\bibitem [{\citenamefont {Altland}\ and\ \citenamefont
  {Simons}(2010)}]{altland_book}%
  \BibitemOpen
  \bibfield  {author} {\bibinfo {author} {\bibfnamefont {A.}~\bibnamefont
  {Altland}}\ and\ \bibinfo {author} {\bibfnamefont {B.}~\bibnamefont
  {Simons}},\ }\href@noop {} {\emph {\bibinfo {title} {Condensed Matter Field
  Theory}}}\ (\bibinfo  {publisher} {Cambridge University Press},\ \bibinfo
  {year} {2010})\BibitemShut {NoStop}%
\bibitem [{\citenamefont {Kamenev}(2011)}]{kamenev_book}%
  \BibitemOpen
  \bibfield  {author} {\bibinfo {author} {\bibfnamefont {A.}~\bibnamefont
  {Kamenev}},\ }\href@noop {} {\emph {\bibinfo {title} {Field Theory of
  Non-Equilibrium Systems}}},\ \bibinfo {edition} {1st}\ ed.\ (\bibinfo
  {publisher} {Cambrdige University Press},\ \bibinfo {year}
  {2011})\BibitemShut {NoStop}%
\bibitem [{\citenamefont {Fr\"{o}hlich}\ and\ \citenamefont
  {Spencer}(1981)}]{frohlich81}%
  \BibitemOpen
  \bibfield  {author} {\bibinfo {author} {\bibfnamefont {J.}~\bibnamefont
  {Fr\"{o}hlich}}\ and\ \bibinfo {author} {\bibfnamefont {T.}~\bibnamefont
  {Spencer}},\ }\href {http://dx.doi.org/10.1007/BF01208273} {\bibfield
  {journal} {\bibinfo  {journal} {Communications in Mathematical Physics}\
  }\textbf {\bibinfo {volume} {81}},\ \bibinfo {pages} {527} (\bibinfo {year}
  {1981})},\ \bibinfo {note} {10.1007/BF01208273}\BibitemShut {NoStop}%
\bibitem [{\citenamefont {Gritsev}\ \emph {et~al.}(2006)\citenamefont
  {Gritsev}, \citenamefont {Altman}, \citenamefont {Demler},\ and\
  \citenamefont {Polkovnikov}}]{gritsev06}%
  \BibitemOpen
  \bibfield  {author} {\bibinfo {author} {\bibfnamefont {V.}~\bibnamefont
  {Gritsev}}, \bibinfo {author} {\bibfnamefont {E.}~\bibnamefont {Altman}},
  \bibinfo {author} {\bibfnamefont {E.}~\bibnamefont {Demler}}, \ and\ \bibinfo
  {author} {\bibfnamefont {A.}~\bibnamefont {Polkovnikov}},\ }\href
  {http://dx.doi.org/10.1038/nphys410} {\bibfield  {journal} {\bibinfo
  {journal} {Nature Physics}\ }\textbf {\bibinfo {volume} {2}},\ \bibinfo
  {pages} {705} (\bibinfo {year} {2006})}\BibitemShut {NoStop}%
\bibitem [{\citenamefont {Mathey}\ and\ \citenamefont
  {Polkovnikov}(2010)}]{mathey10}%
  \BibitemOpen
  \bibfield  {author} {\bibinfo {author} {\bibfnamefont {L.}~\bibnamefont
  {Mathey}}\ and\ \bibinfo {author} {\bibfnamefont {A.}~\bibnamefont
  {Polkovnikov}},\ }\href {\doibase 10.1103/PhysRevA.81.033605} {\bibfield
  {journal} {\bibinfo  {journal} {Phys. Rev. A}\ }\textbf {\bibinfo {volume}
  {81}},\ \bibinfo {pages} {033605} (\bibinfo {year} {2010})}\BibitemShut
  {NoStop}%
\bibitem [{\citenamefont {Zamoldochikov}(1995)}]{zamoldochikov95}%
  \BibitemOpen
  \bibfield  {author} {\bibinfo {author} {\bibfnamefont {A.~B.}\ \bibnamefont
  {Zamoldochikov}},\ }\href@noop {} {\bibfield  {journal} {\bibinfo  {journal}
  {International Journal of Modern Physics A}\ }\textbf {\bibinfo {volume}
  {10}},\ \bibinfo {pages} {1125} (\bibinfo {year} {1995})}\BibitemShut
  {NoStop}%
\bibitem [{\citenamefont {Lukyanov}\ and\ \citenamefont
  {Zamolodchikov}(1997)}]{lukyanov97}%
  \BibitemOpen
  \bibfield  {author} {\bibinfo {author} {\bibfnamefont {S.}~\bibnamefont
  {Lukyanov}}\ and\ \bibinfo {author} {\bibfnamefont {A.}~\bibnamefont
  {Zamolodchikov}},\ }\href {\doibase 10.1016/S0550-3213(97)00123-5} {\bibfield
   {journal} {\bibinfo  {journal} {Nuclear Physics B}\ }\textbf {\bibinfo
  {volume} {493}},\ \bibinfo {pages} {571 } (\bibinfo {year}
  {1997})}\BibitemShut {NoStop}%
\bibitem [{\citenamefont {De~Grandi}, \citenamefont {Gritsev},\ and\
  \citenamefont {Polkovnikov}(2010{\natexlab{b}})}]{degrandi10}%
  \BibitemOpen
  \bibfield  {author} {\bibinfo {author} {\bibfnamefont {C.}~\bibnamefont
  {De~Grandi}}, \bibinfo {author} {\bibfnamefont {V.}~\bibnamefont {Gritsev}},
  \ and\ \bibinfo {author} {\bibfnamefont {A.}~\bibnamefont {Polkovnikov}},\
  }\href {\doibase 10.1103/PhysRevB.81.224301} {\bibfield  {journal} {\bibinfo
  {journal} {Phys. Rev. B}\ }\textbf {\bibinfo {volume} {81}},\ \bibinfo
  {pages} {224301} (\bibinfo {year} {2010}{\natexlab{b}})}\BibitemShut
  {NoStop}%
\bibitem [{\citenamefont {Iucci}\ and\ \citenamefont
  {Cazalilla}(2010)}]{iucci10}%
  \BibitemOpen
  \bibfield  {author} {\bibinfo {author} {\bibfnamefont {A.}~\bibnamefont
  {Iucci}}\ and\ \bibinfo {author} {\bibfnamefont {M.~A.}\ \bibnamefont
  {Cazalilla}},\ }\href {http://stacks.iop.org/1367-2630/12/i=5/a=055019}
  {\bibfield  {journal} {\bibinfo  {journal} {New Journal of Physics}\ }\textbf
  {\bibinfo {volume} {12}},\ \bibinfo {pages} {055019} (\bibinfo {year}
  {2010})}\BibitemShut {NoStop}%
\bibitem [{\citenamefont {White}(1992)}]{white92}%
  \BibitemOpen
  \bibfield  {author} {\bibinfo {author} {\bibfnamefont {S.~R.}\ \bibnamefont
  {White}},\ }\href {\doibase 10.1103/PhysRevLett.69.2863} {\bibfield
  {journal} {\bibinfo  {journal} {Phys. Rev. Lett.}\ }\textbf {\bibinfo
  {volume} {69}},\ \bibinfo {pages} {2863} (\bibinfo {year}
  {1992})}\BibitemShut {NoStop}%
\bibitem [{\citenamefont {Sch\"{o}llwock}(2011)}]{scholl06}%
  \BibitemOpen
  \bibfield  {author} {\bibinfo {author} {\bibfnamefont {U.}~\bibnamefont
  {Sch\"{o}llwock}},\ }\href {\doibase 10.1016/j.aop.2010.09.012} {\bibfield
  {journal} {\bibinfo  {journal} {Annals of Physics}\ }\textbf {\bibinfo
  {volume} {326}},\ \bibinfo {pages} {96 } (\bibinfo {year} {2011})},\ \bibinfo
  {note} {january 2011 Special Issue}\BibitemShut {NoStop}%
\bibitem [{\citenamefont {Bethe}(1931)}]{bethe31}%
  \BibitemOpen
  \bibfield  {author} {\bibinfo {author} {\bibfnamefont {H.}~\bibnamefont
  {Bethe}},\ }\href {http://dx.doi.org/10.1007/BF01341708} {\bibfield
  {journal} {\bibinfo  {journal} {Zeitschrift fur Physik A Hadrons and Nuclei}\
  }\textbf {\bibinfo {volume} {71}},\ \bibinfo {pages} {205} (\bibinfo {year}
  {1931})},\ \bibinfo {note} {10.1007/BF01341708}\BibitemShut {NoStop}%
\bibitem [{\citenamefont {Sirker}\ and\ \citenamefont
  {Bortz}(2006)}]{sirker06}%
  \BibitemOpen
  \bibfield  {author} {\bibinfo {author} {\bibfnamefont {J.}~\bibnamefont
  {Sirker}}\ and\ \bibinfo {author} {\bibfnamefont {M.}~\bibnamefont {Bortz}},\
  }\href {http://stacks.iop.org/1742-5468/2006/i=01/a=P01007} {\bibfield
  {journal} {\bibinfo  {journal} {Journal of Statistical Mechanics: Theory and
  Experiment}\ }\textbf {\bibinfo {volume} {2006}},\ \bibinfo {pages} {P01007}
  (\bibinfo {year} {2006})}\BibitemShut {NoStop}%
\bibitem [{\citenamefont {Vidal}(2004)}]{vidal93}%
  \BibitemOpen
  \bibfield  {author} {\bibinfo {author} {\bibfnamefont {G.}~\bibnamefont
  {Vidal}},\ }\href {\doibase 10.1103/PhysRevLett.93.040502} {\bibfield
  {journal} {\bibinfo  {journal} {Phys. Rev. Lett.}\ }\textbf {\bibinfo
  {volume} {93}},\ \bibinfo {pages} {040502} (\bibinfo {year}
  {2004})}\BibitemShut {NoStop}%
\bibitem [{\citenamefont {Wall}\ and\ \citenamefont {Carr}()}]{openTEBD}%
  \BibitemOpen
  \bibfield  {author} {\bibinfo {author} {\bibfnamefont {M.~L.}\ \bibnamefont
  {Wall}}\ and\ \bibinfo {author} {\bibfnamefont {L.~D.}\ \bibnamefont
  {Carr}},\ }\href {http://physics.mines.edu/downloads/software/tebd/} {\emph
  {\bibinfo {title} {Open Source TEBD (2009)}}}\ (\bibinfo  {publisher}
  {http://physics.mines.edu/downloads/software/tebd/})\BibitemShut {NoStop}%
\bibitem [{\citenamefont {Witthaut}, \citenamefont {Graefe},\ and\
  \citenamefont {Korsch}(2006)}]{witthaut06}%
  \BibitemOpen
  \bibfield  {author} {\bibinfo {author} {\bibfnamefont {D.}~\bibnamefont
  {Witthaut}}, \bibinfo {author} {\bibfnamefont {E.~M.}\ \bibnamefont
  {Graefe}}, \ and\ \bibinfo {author} {\bibfnamefont {H.~J.}\ \bibnamefont
  {Korsch}},\ }\href {\doibase 10.1103/PhysRevA.73.063609} {\bibfield
  {journal} {\bibinfo  {journal} {Phys. Rev. A}\ }\textbf {\bibinfo {volume}
  {73}},\ \bibinfo {pages} {063609} (\bibinfo {year} {2006})}\BibitemShut
  {NoStop}%
\bibitem [{\citenamefont {Schumm}\ \emph {et~al.}(2005)\citenamefont {Schumm},
  \citenamefont {Hofferberth}, \citenamefont {Andersson}, \citenamefont
  {Wildermuth}, \citenamefont {Groth}, \citenamefont {Bar-Joseph},
  \citenamefont {Schmiedmayer},\ and\ \citenamefont {Kruger}}]{schumm05}%
  \BibitemOpen
  \bibfield  {author} {\bibinfo {author} {\bibfnamefont {T.}~\bibnamefont
  {Schumm}}, \bibinfo {author} {\bibfnamefont {S.}~\bibnamefont {Hofferberth}},
  \bibinfo {author} {\bibfnamefont {L.~M.}\ \bibnamefont {Andersson}}, \bibinfo
  {author} {\bibfnamefont {S.}~\bibnamefont {Wildermuth}}, \bibinfo {author}
  {\bibfnamefont {S.}~\bibnamefont {Groth}}, \bibinfo {author} {\bibfnamefont
  {I.}~\bibnamefont {Bar-Joseph}}, \bibinfo {author} {\bibfnamefont
  {J.}~\bibnamefont {Schmiedmayer}}, \ and\ \bibinfo {author} {\bibfnamefont
  {P.}~\bibnamefont {Kruger}},\ }\href {http://dx.doi.org/10.1038/nphys125}
  {\bibfield  {journal} {\bibinfo  {journal} {Nature Physics}\ }\textbf
  {\bibinfo {volume} {1}},\ \bibinfo {pages} {57} (\bibinfo {year}
  {2005})}\BibitemShut {NoStop}%
\bibitem [{\citenamefont {Gring}\ \emph {et~al.}(2012)\citenamefont {Gring},
  \citenamefont {Kuhnert}, \citenamefont {Langen}, \citenamefont {Kitagawa},
  \citenamefont {Rauer}, \citenamefont {Schreitl}, \citenamefont {Mazets},
  \citenamefont {Smith}, \citenamefont {Demler},\ and\ \citenamefont
  {Schmiedmayer}}]{gring12}%
  \BibitemOpen
  \bibfield  {author} {\bibinfo {author} {\bibfnamefont {M.}~\bibnamefont
  {Gring}}, \bibinfo {author} {\bibfnamefont {M.}~\bibnamefont {Kuhnert}},
  \bibinfo {author} {\bibfnamefont {T.}~\bibnamefont {Langen}}, \bibinfo
  {author} {\bibfnamefont {T.}~\bibnamefont {Kitagawa}}, \bibinfo {author}
  {\bibfnamefont {B.}~\bibnamefont {Rauer}}, \bibinfo {author} {\bibfnamefont
  {M.}~\bibnamefont {Schreitl}}, \bibinfo {author} {\bibfnamefont
  {I.}~\bibnamefont {Mazets}}, \bibinfo {author} {\bibfnamefont {D.~A.}\
  \bibnamefont {Smith}}, \bibinfo {author} {\bibfnamefont {E.}~\bibnamefont
  {Demler}}, \ and\ \bibinfo {author} {\bibfnamefont {J.}~\bibnamefont
  {Schmiedmayer}},\ }\href {\doibase 10.1126/science.1224953} {\bibfield
  {journal} {\bibinfo  {journal} {Science}\ }\textbf {\bibinfo {volume}
  {337}},\ \bibinfo {pages} {1318} (\bibinfo {year} {2012})}\BibitemShut
  {NoStop}%
\bibitem [{\citenamefont {Mitra}\ and\ \citenamefont
  {Giamarchi}(2011)}]{mitra11}%
  \BibitemOpen
  \bibfield  {author} {\bibinfo {author} {\bibfnamefont {A.}~\bibnamefont
  {Mitra}}\ and\ \bibinfo {author} {\bibfnamefont {T.}~\bibnamefont
  {Giamarchi}},\ }\href {\doibase 10.1103/PhysRevLett.107.150602} {\bibfield
  {journal} {\bibinfo  {journal} {Phys. Rev. Lett.}\ }\textbf {\bibinfo
  {volume} {107}},\ \bibinfo {pages} {150602} (\bibinfo {year}
  {2011})}\BibitemShut {NoStop}%
\bibitem [{\citenamefont {Dalla~Torre}\ \emph {et~al.}(2012)\citenamefont
  {Dalla~Torre}, \citenamefont {Demler}, \citenamefont {Giamarchi},\ and\
  \citenamefont {Altman}}]{dallatorre11}%
  \BibitemOpen
  \bibfield  {author} {\bibinfo {author} {\bibfnamefont {E.~G.}\ \bibnamefont
  {Dalla~Torre}}, \bibinfo {author} {\bibfnamefont {E.}~\bibnamefont {Demler}},
  \bibinfo {author} {\bibfnamefont {T.}~\bibnamefont {Giamarchi}}, \ and\
  \bibinfo {author} {\bibfnamefont {E.}~\bibnamefont {Altman}},\ }\href
  {\doibase 10.1103/PhysRevB.85.184302} {\bibfield  {journal} {\bibinfo
  {journal} {Phys. Rev. B}\ }\textbf {\bibinfo {volume} {85}},\ \bibinfo
  {pages} {184302} (\bibinfo {year} {2012})}\BibitemShut {NoStop}%
\bibitem [{\citenamefont {Mitra}\ and\ \citenamefont
  {Giamarchi}(2012)}]{mitra12}%
  \BibitemOpen
  \bibfield  {author} {\bibinfo {author} {\bibfnamefont {A.}~\bibnamefont
  {Mitra}}\ and\ \bibinfo {author} {\bibfnamefont {T.}~\bibnamefont
  {Giamarchi}},\ }\href {\doibase 10.1103/PhysRevB.85.075117} {\bibfield
  {journal} {\bibinfo  {journal} {Phys. Rev. B}\ }\textbf {\bibinfo {volume}
  {85}},\ \bibinfo {pages} {075117} (\bibinfo {year} {2012})}\BibitemShut
  {NoStop}%
\bibitem [{\citenamefont {Mitra}(2012)}]{mitra12b}%
  \BibitemOpen
  \bibfield  {author} {\bibinfo {author} {\bibfnamefont {A.}~\bibnamefont
  {Mitra}},\ }\href {\doibase 10.1103/PhysRevLett.109.260601} {\bibfield
  {journal} {\bibinfo  {journal} {Phys. Rev. Lett.}\ }\textbf {\bibinfo
  {volume} {109}},\ \bibinfo {pages} {260601} (\bibinfo {year}
  {2012})}\BibitemShut {NoStop}%
\bibitem [{\citenamefont {Luther}\ and\ \citenamefont
  {Peschel}(1974)}]{luther74}%
  \BibitemOpen
  \bibfield  {author} {\bibinfo {author} {\bibfnamefont {A.}~\bibnamefont
  {Luther}}\ and\ \bibinfo {author} {\bibfnamefont {I.}~\bibnamefont
  {Peschel}},\ }\href@noop {} {\bibfield  {journal} {\bibinfo  {journal} {Phys.
  Rev. B}\ }\textbf {\bibinfo {volume} {9}},\ \bibinfo {pages} {2911} (\bibinfo
  {year} {1974})}\BibitemShut {NoStop}%
\bibitem [{\citenamefont {Bistritzer}\ and\ \citenamefont
  {Altman}(2007)}]{bistritzer07}%
  \BibitemOpen
  \bibfield  {author} {\bibinfo {author} {\bibfnamefont {R.}~\bibnamefont
  {Bistritzer}}\ and\ \bibinfo {author} {\bibfnamefont {E.}~\bibnamefont
  {Altman}},\ }\href {\doibase 10.1073/pnas.0608910104} {\bibfield  {journal}
  {\bibinfo  {journal} {Proceedings of the National Academy of Sciences}\
  }\textbf {\bibinfo {volume} {104}},\ \bibinfo {pages} {9955} (\bibinfo {year}
  {2007})}\BibitemShut {NoStop}%
\bibitem [{\citenamefont {Barmettler}\ \emph {et~al.}(2010)\citenamefont
  {Barmettler}, \citenamefont {Punk}, \citenamefont {Gritsev}, \citenamefont
  {Demler},\ and\ \citenamefont {Altman}}]{barmettler09}%
  \BibitemOpen
  \bibfield  {author} {\bibinfo {author} {\bibfnamefont {P.}~\bibnamefont
  {Barmettler}}, \bibinfo {author} {\bibfnamefont {M.}~\bibnamefont {Punk}},
  \bibinfo {author} {\bibfnamefont {V.}~\bibnamefont {Gritsev}}, \bibinfo
  {author} {\bibfnamefont {E.}~\bibnamefont {Demler}}, \ and\ \bibinfo {author}
  {\bibfnamefont {E.}~\bibnamefont {Altman}},\ }\href
  {http://stacks.iop.org/1367-2630/12/i=5/a=055017} {\bibfield  {journal}
  {\bibinfo  {journal} {New Journal of Physics}\ }\textbf {\bibinfo {volume}
  {12}},\ \bibinfo {pages} {055017} (\bibinfo {year} {2010})}\BibitemShut
  {NoStop}%
\end{thebibliography}%

\end{document}